\documentclass[a4paper,11pt]{article}
\pdfoutput=1 
\usepackage{jcappub} 
\usepackage[T1]{fontenc}

\newcommand{\gsim}{\raisebox{-0.13cm}{~\shortstack{$>$ \\[-0.07cm]
      $\sim$}}~}
\newcommand{\lsim}{\raisebox{-0.13cm}{~\shortstack{$<$ \\[-0.07cm]
      $\sim$}}~}

\usepackage{subfig}
\usepackage{comment}
\usepackage{graphicx}
\usepackage{amsmath}
\usepackage{amssymb}
\usepackage{lipsum}
\usepackage{multirow}
\usepackage{xcolor}
\usepackage{hyperref}
\usepackage{multicol}
\usepackage{float}
\usepackage{soul}
\usepackage[normalem]{ulem}
\usepackage{aas_macros}

\defcitealias{Mead_2021}{M20}
\defcitealias{Takahashi_2012}{TB}

\title{\centering{\boldmath{\fontsize{19}{11}\selectfont DEMNUni: cross-correlating the nonlinear ISWRS effect with CMB-lensing and galaxies\\ \vspace{-0.5cm} in the presence of massive neutrinos}}}

\author[a,b,c]{Viviana Cuozzo}
\author[d]{Carmelita Carbone}
\author[e]{Matteo Calabrese}
\author[d,a]{Elisabetta Carella}
\author[b,c]{and Marina Migliaccio}

\affiliation[a]{Dipartimento di Fisica ``Aldo Pontremoli'', Universit\`{a} degli Studi di Milano, via Celoria 16, I-20133 Milano, Italy}
\affiliation[b]{Università degli Studi di Roma Tor Vergata, via della Ricerca Scientifica 1, I-00133 Roma, Italy}
\affiliation[c]{INFN, Sezione di Roma 2, Università degli Studi di Roma Tor Vergata, via della Ricerca Scientifica 1, I-00133 Roma, Italy }
\affiliation[d]{INAF -- Istituto di Astrofisica Spaziale e Fisica cosmica di Milano (IASF-MI), via Alfonso Corti 12, I-20133 Milano, Italy}
\affiliation[e]{Astronomical Observatory of the Autonomous Region of the Aosta Valley (OAVdA), Loc. Lignan 39, I-11020, Nus (Aosta Valley), Italy}

\emailAdd{vcuozzo@roma2.infn.it}

\abstract{We present an analytical modelling of the angular cross-correlations between the Integrated Sachs Wolfe--Rees Sciama (ISWRS) effect and large-scale structure tracers in the presence of massive neutrinos. Our method has been validated against large N-body simulations with a massive neutrino particle component, namely the DEMNUni suite. We investigate the impact of different neutrino masses on the cross-correlations between the ISWRS effect and both the galaxy clustering and the lensing of the Cosmic Microwave Background (CMB). We also test the ability of current nonlinear matter power spectrum modellings to reproduce neutrino effects in such cross-correlations. We show that the multipole position of a characteristic sign inversion in the cross-spectra, due to nonlinear effects, is strongly related to the total neutrino mass $M_\nu$ and depends almost linearly on it. While these nonlinear cross-correlation signals may not be able alone to constrain the neutrino mass, our approach paves the way to the detection of such cross-spectra on small scales for their exploitation in combination with main probes from future galaxy surveys and CMB experiments.
}

\begin{document}

\maketitle 

\section{Introduction}
\label{sec:intro}
Since its discovery in 1965~\cite{Penzias_1965}, the CMB has provided the most significant evidence for the Big Bang Theory and the study of its anisotropies has given us a unique window onto the primordial Universe. Measurements from WMAP~\cite{Bennett_2003,Spergel_2003,Komatsu_2011} and  Planck~\cite{PlanckXVI_2014, PlanckXIII_2016, PlanckVI_2020} have consistently been in line with the predictions of the $\Lambda$CDM model, the so-called Standard Cosmological Model, and have been fundamental in inferring constraints on the cosmological parameters that characterise it. However, there are still many open questions in cosmology, and several tensions exist in the Standard Cosmological Model. One of the most investigated topics today is the nature of Dark Energy (DE), which appears to dominate the energy density of the Universe at late times (see e.g. ~\cite{Linder_2003}), counteracting the gravitational collapse and inducing a background accelerated expansion. Another example is the value of the sum of the three neutrino masses, $M_{\nu} = \sum m_{\nu}$, whose best bounds come from cosmological data~\cite{Lesgourgues_2012, Arcidiacono_2020, PlanckVI_2020}. 

Among several cosmological probes, both these issues can be addressed also by measuring and analysing the late Integrated Sachs Wolfe (ISW) effect~\cite{Sachs_1967}, which is the temperature variation that CMB photons experience when they propagate in a time-varying gravitational potential, and the Rees Sciama (RS) effect~\cite{Rees_1968}, which is its nonlinear counterpart. DE is mainly responsible for the combined ISWRS signal \cite{Ferraro_2015}, but it is not the unique source of such effect. The presence of massive neutrinos induces a slow decay of the gravitational potential~\cite{Bond_1980} that generates ISWRS even in the absence of a background expansion. Therefore, a full reconstruction of the ISWRS effect would provide new information on the physics of neutrinos and of DE, allowing to help constraining the DE Equation of State (EoS) and the total neutrino mass. However, in this respect there has not been much progress as this signal is extremely faint compared to CMB primary anisotropies. Moreover, these measurements are limited in the linear regime by cosmic variance and in the nonlinear regime by the lack of data. The only way to reconstruct the ISWRS is to use its cross-correlation with the Large Scale Structure (LSS), in order to exploit the tracers that follow the dark matter field and the variation of the gravitational potential produced by DE and massive neutrinos~\cite{Crittenden_1996, Kneissl_1997, Boughn_1998, Boughn_2001, Boughn_2004, Giannantonio_2008, Lesgourgues_2008, Douspis_2008}.  

In this work, we focus on the cross-correlations of the ISWRS with the CMB-lensing (CMBL) potential and the galaxy distribution. As the RS effect anti-correlates with both of them~\cite{Cai_2009, Nishizawa_2014, Carbone_2016}, these cross-spectra are characterised by the presence of a sign inversion that occurs when passing from the linear (ISW) to the nonlinear (RS) regime. Moreover, the position of the sign inversion, which indicates the appearance of nonlinearities, shifts towards smaller cosmological scales due to the suppression of the matter power spectrum resulting from the presence of massive neutrinos~\cite{Cai_2009, Carbone_2016}. 

We develop an analytical method to produce the ISWRS spectra using the nonlinear modelling of the matter power spectrum implemented in the Boltzmann solver code \texttt{CAMB}\footnote{\url{https://CAMB.readthedocs.io/en/latest/}}~\cite{Lewis_2011}. Given the large redshift extension investigated in this work \cite{Parimbelli_2022}, we consider only two of the revised \texttt{Halofit}~\cite{Smith_2003, Cooray_2002} models implemented: the Takahashi2012+Bird2014 (hereafter \citetalias{Takahashi_2012}) model~\cite{Takahashi_2012}, extended to massive neutrino cosmologies by including Bird correction\footnote{As reported in the Readme of the \texttt{CAMB} webpage, on March 2014 modified massive neutrino parameters were implemented in the nonlinear fitting of the total matter power spectrum to improve the accuracy of the updated \texttt{Halofit} version from~\cite{Takahashi_2012}. These fitting parameters, accounting for nonlinear corrections in the presence of massive neutrinos, are different from the ones
implemented by~\cite{Bird_2012} in the original HF version from Smith et al. (2003) \cite{Smith_2003}.} \cite{Bird_2012}; the Mead2020 (hereafter \citetalias{Mead_2021}) model~\cite{Mead_2021}, which is an improvement over the Mead2016 model~\cite{Mead_2015} concerning the treatment of Baryon Acoustic Oscillations (BAO) nonlinear damping. 
We then validate this analytical approach by comparing it with signals extracted from the ``Dark Energy and Massive Neutrino Universe'' (DEMNUni) N-body numerical simulations~\cite{Carbone_2016}.

These studies are particularly relevant in light of ongoing and forthcoming galaxy surveys, such as \textit{Euclid}~\cite{Laureijs_2011, EuclidVII_2020, EuclidI_2022} and SKA~\cite{SKA_cosmology, SKA_cross2, SKA_cross, SKA_LS, SKA_WL, SKA_fundamentalPhys,SKA_Cosmo_Fund}, and CMB surveys, such as CMB-S4~\cite{Abazajian_2016, Abazajian_2022}, which aim to achieve measurements of small scales with high precision and accuracy, such that the possibility of detecting the RS effect could become concrete~\cite{Ferraro_2022}. 

This paper is organised as follows. In Section~\ref{sec:theory} we recap the theoretical framework of the cross-correlations of the ISWRS signal with the CMBL potential and the galaxy distribution in the presence of massive neutrinos. In Section~\ref{sec:simulations} we present how we model ISWRS nonlinearities using measurements from the DEMNUni simulations. In Section~\ref{sec:XC-TP} we test the analytical method for the cross-correlation between the ISWRS signal and the CMBL potential against the DEMNUni simulated cross-spectrum. In Section~\ref{sec:XC-TG} we validate the cross-correlation between the ISWRS signal and the galaxy distribution against the cross-spectrum measured from the DEMNUni mock maps. Moreover, we model the ISWRS--galaxy cross-spectrum for different redshift ranges. In Section~\ref{sec:detection}, we evaluate the predictability of $M_{\nu}$ via the estimation of the sign inversion position in both the analysed cross-correlations. Finally, in Section~\ref{sec:conclusion} we draw our conclusions.

\section{The Integrated Sachs Wolfe and the Rees Sciama effect}
\label{sec:theory}
When CMB photons pass through a time-varying gravitational potential well along their path from the last scattering surface to us, they 
undergo a variation in energy that translates into a variation in temperature known as the Integrated Sachs Wolfe effect:
\begin{equation}
    \frac{\Delta T_{\rm ISW}}{T_{0}} (\hat{\boldsymbol{n}}) = \frac{2}{c^{2}} \int^{t_{0}}_{t_{ls}} {\rm d}t \, \dot{\Phi}(\hat{\boldsymbol{n}}, \chi, t) \,,
\end{equation}
where $\hat{\boldsymbol{n}}$ is a unit direction vector on the sphere, $T_{0}$ is today ($t_{0}$) CMB temperature, $t_{ls}$ is the time at the last scattering surface, $\chi$ the comoving distance and $\dot\Phi$ is the time derivative of the gravitational potential.
This effect is mainly due to the presence of DE that induces an accelerated background expansion of the Universe and counteracts the gravitational potential $\Phi$ inducing a not vanishing $\dot{\Phi}$~\cite{Giannantonio_2008, Giannantonio_2012, Watson_2014, Naidoo_2021}. The nonlinear growth of density perturbations produces additional temperature perturbations, leading to the RS effect. The RS contribution to temperature fluctuations is much more slow over time than the late ISW one, which means that at high redshifts RS predominates over the late ISW~\cite{Cai_2010}. Consequently, we are able to observe $\dot{\Phi}\neq$ 0 even when the DE effect becomes negligible.
Moreover, the presence of massive neutrinos induces a not vanishing derivative of the gravitational potential even during matter domination era.
This is because, after becoming non-relativistic, neutrinos free-stream with large thermal velocities that suppress the growth of neutrino density perturbations on scales smaller than the so-called ``free-streaming length''~\cite{Lesgourgues_2008,Lesgourgues_2012}:
\begin{equation}
    \lambda_{\rm FS}(z, m_{\nu}) \simeq 8.1 \frac{H_{0}(1+z)}{H(z)} \frac{1\text{ eV}}{ m_{\nu}}\, h^{-1}\text{Mpc} \,,
\end{equation}
where $m_{\nu}$ is the mass of the single neutrino species (i.e. $\nu_{e}, \nu_{\mu} \text{ or } \nu_{\tau}$), $H(z)$ is the Hubble parameter as a function of the redshift $z$ and $H_{0} \equiv H(z=0)$ is the Hubble constant.  Moreover, because of the gravitational back reaction effects, the evolution of cold dark matter (CDM) and baryon densities is affected as well by the presence of neutrinos, and the total matter power spectrum is suppressed at scales $\lambda \ll \lambda_{\rm FS}$~\cite{Rossi_2014}. Consequently, on small cosmological scales the free-streaming of neutrinos induces a slow decay of the gravitational potential, acting during both the matter and the DE dominated eras, and this effect depends on the total neutrino mass $M_{\nu} = \sum m_{\nu}$.  The more neutrinos are massive, the more the matter power spectrum will be suppressed with respect to the massless neutrino case, and therefore nonlinearities will appear on smaller scales. 

One of the effects of massive neutrinos, which has been poorly investigated so far, is the shift they induce in the sign inversion that characterises the cross-power spectra between the ISWRS signal and the CMBL potential ($P_{\dot\Phi\Phi}$), and between the ISWRS signal and the galaxy distribution ($P_{\dot\Phi\delta}$). These sign inversions appear because of the anti-correlation between the RS and both the CMBL potential and the galaxy distribution, that depend on the gravitational potential $\Phi$. While on large scales (i.e. the linear regime) the gravitational potential decays because of the Universe expansion ($\Phi<0$, $\dot\Phi<0$), on small scales (i.e. the nonlinear regime) $\Phi$ grows because of nonlinear structures formation ($\Phi<0$, $\dot\Phi>0$). Consequently, the time derivative of the gravitational potential associated to the ISWRS signal is negative in the linear regime and positive in the nonlinear one. The net result is that $\langle \Phi\dot{\Phi}\rangle > 0$ in the late ISW regime, producing a positive correlation, and $\langle \Phi\dot{\Phi}\rangle < 0$ in the RS regime, producing an anti-correlation~\cite{Nishizawa_2014, Nishizawa_2008}. 

Following the example of~\cite{Seljak_1996, Cai_2009, Smith_2009}, we report below the dimensionless and scaled (i.e. divided by ${3}/{2} [H_0/(ck)]^2 \Omega_{m}$) forms of $P_{\dot{\Phi}\Phi}$ and $P_{\dot{\Phi}\delta}$ in  Limber approximation~\cite{Smith_2009, Lesgourgues_2008}: 
\begin{equation}
    \Delta^{2}_{\dot\Phi\Phi} = \frac{4\pi}{(2\pi)^{3}} \frac{k^{3}P_{\dot\Phi\Phi}(k,z)}{[F(k)H(z)/ca(z)]}\, \text{, with } P_{\dot\Phi\Phi}(k,z) = - \frac{1}{2} \frac{H(z)}{a(z)} [F(k)]^{2} \Big(\partial_{z}\frac{P_{\delta\delta}(k, z)}{a^{2}(z)} \Big) \,,
 \label{eq:delta_ptp}
\end{equation}
\begin{equation}
 \label{eq:delta_ptg}
     \Delta^{2}_{\dot\Phi\delta} = \frac{4\pi}{(2\pi)^{3}} \frac{k^{3}P_{\dot\Phi\delta}(k,z)}{[F(k)H(z)/ca(z)]} \, \text{, with } P_{\dot\Phi\delta}(k,z) = \frac{1}{2} H(z)F(k) \Big(\partial_{z}\frac{P_{\delta\delta}(k, z)}{a^{2}(z)} \Big) \,,
 \end{equation}
where $a(z)$ is the scale factor, $P_{\delta\delta}$ is the matter power spectrum, $k$ is the wavenumber, and
\begin{equation*}
    F(k) = \frac{3 \Omega_{m} H_{0}^{2}}{2 c^{2} k^{2}}\,,
\end{equation*}
with $\Omega_{m}$ the matter density parameter and $c$ the speed of light.

\begin{figure*}[!ht]
 \centering
\includegraphics[width=0.95\textwidth]{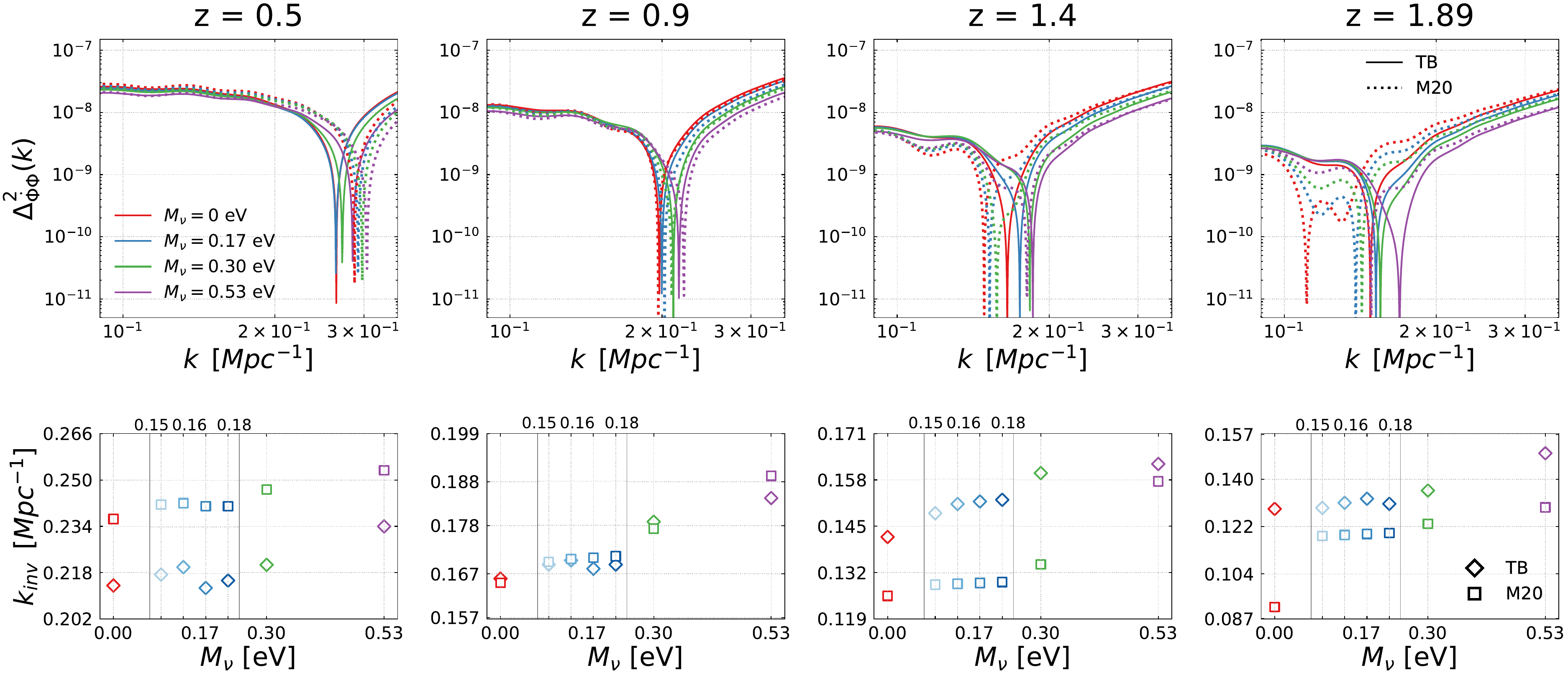} 
\includegraphics[width=0.95\textwidth]{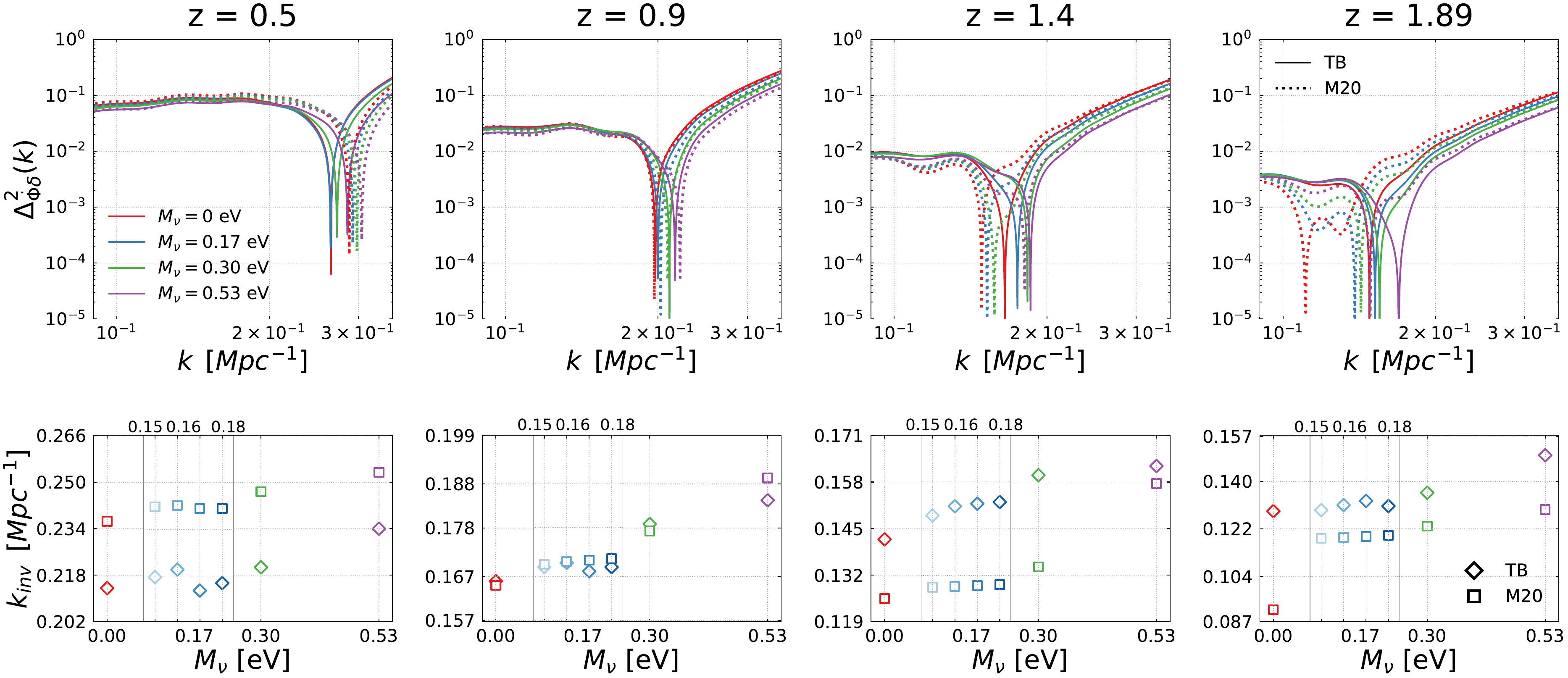}
\caption{\textit{First row}: Absolute values of the dimensionless and scaled form of the ISWRS--CMBL potential cross-spectra as a function of the wavenumber $k$ at $z=0.5, 0.9, 1.4, 1.89$ for the four neutrino total masses considered ($M_{\nu}=0,0.17,0.30, 0.53$ eV as red, blue, green and purple lines, respectively), obtained via the \citetalias{Takahashi_2012} model (solid lines) and the \citetalias{Mead_2021} model (dotted lines). \textit{Second row}: Measured $k_{inv}$ as a function of $M_{\nu}$ for the \citetalias{Takahashi_2012} (diamonds) and the \citetalias{Mead_2021} (squares) models.
\textit{Third row}: Absolute values of the dimensionless and scaled form of the ISWRS--galaxy distribution cross-spectra as a function of the wavenumber $k$ at $z=0.5, 0.9, 1.4, 1.89$ for the four neutrino masses considered ($M_{\nu}=0,0.17,0.30, 0.53$ eV as red, blue, green and purple lines, respectively), obtained via the \citetalias{Takahashi_2012} model (solid lines) and the \citetalias{Mead_2021} model (dotted lines). \textit{Fourth row}: Measured $k_{inv}$ as a function of $ M_{\nu}$ for the \citetalias{Takahashi_2012} (diamonds) and the \citetalias{Mead_2021} (squares) models.}
\label{spectra_comp1_comp2}
\end{figure*}

We compute each term of Equations~\eqref{eq:delta_ptp} and \eqref{eq:delta_ptg} with \texttt{CAMB} and use both the nonlinear modellings chosen for the matter power spectrum (i.e. \citetalias{Takahashi_2012} and \citetalias{Mead_2021}). The absolute values of the results are shown in the panels of the first and third rows of Figure~\ref{spectra_comp1_comp2}, for both the \citetalias{Takahashi_2012} and \citetalias{Mead_2021} nonlinear modellings, at different redshift values, for the four total neutrino masses considered in this work.

The panels in the first and third rows of Figure~\ref{spectra_comp1_comp2} show how the presence of nonlinearities varies with time. As already mentioned, the more we travel back in time, the more the RS effect will dominate over the late ISW. Considering that at large redshifts the RS traces not only nonlinearities at small scales, but even the production of filamentary structures (visible at scales of few hundreds Mpc)~\cite{Cai_2010}, we see the shifts of the sign inversion moving towards smaller $k$ (i.e. larger cosmological scales) from $z=0.5$ to $z=1.89$.  Following \cite{Smith_2009}, we verified that, in the redshift range of our interest, the linear growth rate  of matter perturbations, $f \equiv d\ln(D) / d \ln(a)$ (where $D$ is the so-called linear growth factor) is $f  > 0.5$, which is the condition expected to be necessary for the presence of the sign inversions measured in the cross power spectra.

In the panels of the second and fourth rows in Figure~\ref{spectra_comp1_comp2}, we present the trend of $k_{inv}$ as a function of $M_{\nu}$ for both the nonlinear models tested. We observe the expected effect: the more massive neutrinos are, the more nonlinearities (and therefore, the sign inversion) move toward smaller cosmological scales (i.e. larger $k$). The only exception, for both $\Delta^{2}_{\dot\Phi \Phi}$ and $\Delta^{2}_{\dot\Phi\delta}$, is the trend obtained for $M_{\nu} = 0.17$ eV when using the \citetalias{Takahashi_2012} model at $z = 0.5$, where the power spectra show the sign inversion at scales slightly larger than the massless neutrino case. We additionally include the results for $M_{\nu}$ values slightly smaller and larger than $0.17$ eV. On the one hand, we observe an increase of $k_{inv}$ from $M_{\nu} = 0$ eV to $M_{\nu} = 0.15, 0.16$ eV, on the other hand, we confirm a decrease corresponding to $M_{\nu} = 0.17$ eV, followed by a new increase for $M_{\nu} = 0.18$ eV. This effect is probably due to inaccuracies in the \citetalias{Takahashi_2012} nonlinear modelling.

The similarity of the results shown in the panels of the second and fourth rows is due to the fact that the position of $k_{inv}$ as a function of $M_{\nu}$ strictly depends on the time derivative of $P_{\delta\delta}$, that appears and is the same in both Equations~\eqref{eq:delta_ptp}-\eqref{eq:delta_ptg}. All the other terms of these two equations are responsible for the differences in the amplitude alone of the analysed cross-spectra, as it is visible in the panels of the first and third rows.

These results highlight the different behaviour of the two models. At small redshifts the \citetalias{Takahashi_2012} model predicts an higher level of nonlinearities than the \citetalias{Mead_2021} model. Indeed, in the panels of Figure~\ref{spectra_comp1_comp2} corresponding to $z=0.5$, the \citetalias{Takahashi_2012} $k_{inv}$  are smaller than the \citetalias{Mead_2021} ones. However, it is easy to notice that at $z=0.9$ and $z=1.4$ there is a switch of trend between \citetalias{Takahashi_2012} and \citetalias{Mead_2021}, and at $z=1.89$ the \citetalias{Mead_2021} model predicts an higher level of nonlinearities with respect to \citetalias{Takahashi_2012}. Such behaviours strictly depend on the approximations adopted for the implementation of the two \texttt{Halofit} models tested, which however do not aim at reproducing correctly the time derivative of the gravitational potential sourcing the ISWRS effect.

The presence of massive neutrinos is expected to affect similarly also the angular power-spectra of both these cross-correlations, that actually depend on the time derivative of the matter power spectrum $P_{\delta\delta}$, and can be computed via~\cite{Mangilli_2009}:
\begin{equation}
    C_{\ell}^{\dot\Phi\Phi} = \frac{1}{2} \Big(\frac{3 \Omega_{m} H_{0}^{2}}{2c^{2}k^{2}}\Big)^{2} \int_{z_{\rm min}}^{z_{\rm max}} {\rm d}z\, \frac{\chi(z_{ls}) - \chi(z)}{\chi(z_{ls})\chi^{3}(z)} \Big[ \partial_{z} \frac{P_{\delta\delta}(k,z)}{a^{2}(z)}\Big] \,,
     \label{ctp}
\end{equation}
where $z_{ls} = 1100$, and~\cite{Lesgourgues_2008, Smith_2009, Ferraro_2022}:
\begin{equation}
    C_{\ell}^{\dot\Phi g} = \frac{3\Omega_{m}H_{0}^{2}}{2c^{3}(\ell + 1/2)^{2}}  \int_{z_{\rm min}}^{z_{\rm max}} {\rm d}z\, n(z)b(z)H(z)a(z)\Big[\partial_{z}\frac{P_{\delta\delta}(k,z)}{a^{2}(z)}\Big] \,, 
    \label{ctg}
\end{equation} 
where $n(z)$ and $b(z)$ are the galaxy selection function and the galaxy bias, respectively. 

The integral limits used in this work are: $z_{\rm min} = 0.02$ and $z_{\rm max} = 1.89$. 
This is because we use Equations~\eqref{ctp} and \eqref{ctg}  to compute the theoretical predictions that are compared against the cross-spectra measured from the DEMNUni maps, that cover the redshift range $[0.02, 1.89]$ (see Section~\ref{sec:simulations}). 

Apart from $n(z)$ and $b(z)$, all the ingredients of Equations~\eqref{ctp} and \eqref{ctg} have been obtained via \texttt{CAMB}.

\section{Nonlinear modelling with N-body simulations}
\label{sec:simulations}
The choice to investigate the nonlinear regime requires the use of large N-body simulations for validation tests. For this work, we use the ``Dark Energy and Massive Neutrino Universe'' (DEMNUni)~\cite{Carbone_2016} N-body cosmological simulations to model at small scales the ISWRS signal and their cross-correlations with the CMBL and the galaxy distributions. 

\subsection{The Dark Energy and Massive Neutrino Universe simulations}
The DEMNUni simulations have been produced with the aim of investigating large-scale structures in the presence of massive neutrinos and dynamical dark energy (DDE), and they were conceived for the nonlinear analysis and modelling of different probes, including dark matter, halo, and galaxy clustering~\cite{Castorina_2015,Moresco_2017,Zennaro_2018,Ruggeri_2018,Bel_2019,Parimbelli_2021,Parimbelli_2022, Baratta_2022, Guidi_2022, Gouyou_Beauchamps_2023, Carella_in_prep}, weak lensing, CMBL, SZ and late ISW effects~\cite{Roncarelli_2015,Carbone_2016,Fabbian_2018, Beatriz_2023}, cosmic void statistics~\cite{Kreisch_2019,Schuster_2019,Verza_2019,Verza_2022a,Verza_2022b}, and cross-correlations among these probes~\cite{Vielzeuf_2023}.
To this end, they combine a good mass resolution with a large volume to include perturbations both at large and small scales. In fact, these simulations follow the evolution of 2048$^3$ cold dark matter (CDM) and, when present, 2048$^3$ neutrino particles in a box of side $L = 2 \, h^{-1} {\rm Gpc}$. The fundamental frequency of the comoving particle snapshot is, therefore, $k_{\rm F} \approx 3\times 10^{-3} \ h\text{ Mp}c^{-1}$, while the chosen softening length is 20 kpc/$h$. The simulations are initialised at $z_{\rm ini}=99$ with Zel'dovich initial conditions. The initial power spectrum is rescaled to the initial redshift via the rescaling method developed in Ref.~\cite{Zennaro_2017}. Initial conditions are then generated with a modified version of the \texttt{N-GenIC} software, assuming Rayleigh random amplitudes and uniform random phases. 
The DEMNUni simulations have been performed using the tree particle mesh-smoothed particle hydrodynamics (TreePM-SPH) code \texttt{GADGET-3}, an improved version of the code described in~\cite{Springel_2005}, specifically
modified in~\cite{Viel_2010} to account for the presence of massive neutrinos. This version of \texttt{GADGET-3} follows the evolution of CDM and neutrino particles, treating them as two distinct collisionless fluids. Given the large amount of memory required by the simulations, baryon physics is not included. This choice, however, does not affect the results of this work~\cite{Daalen_2011, Bird_2012, Castorina_2015}. The DEMNUni suite is composed by a total of sixteen simulations accounting for different combinations of the total neutrino mass $M_{\nu}$ and the DE EoS. In this project, we have worked with four of them, characterised by a baseline Planck 2013 cosmology~\cite{PlanckXVI_2014},  namely a flat $\Lambda$CDM model with $\tau = 0.0925$, $n_{s} =0.96$ and $A_{s}=2.13 \times 10^{-9}$, generalised to a $\nu\Lambda$CDM by varying only the sum of the neutrino masses over the values $M_{\nu} = 0,0.17,0.30,0.53\text{ eV}$ (whence the corresponding values of $\Omega_{\nu}$ and $\Omega_{cdm}$,  keeping fixed $\Omega_{m}$ = 0.32 and $\Omega_{b}$ = 0.05), considering a degenerate neutrino mass hierarchy. As aforementioned, in this work we use maps that cover a range of redshift that goes from $z_{\rm min} = 0.02$ to $z_{\rm max} = 1.89$.
For each simulation, 63 outputs have been produced, logarithmically equispaced in the scale factor $a = 1/(1 + z)$, in the redshift interval $z = 0-99$, $49$ of which lay between $z = 0$ and $z = 10$. For each of the $63$ output times, it has been produced on-the-fly one particle snapshot, composed of both CDM and neutrino particles, one 3D grid of the gravitational potential, $\Phi$, and one 3D grid of its time derivative, $\dot\Phi$, with a mesh of dimension $4096^{3}$ that covers a comoving volume of  $(2 \, h^{-1} {\rm Gpc} )^{3}$.

The map-making procedure has been developed taking into account the standard pixelisation approach introduced by the Hierarchical Equal-Area isoLatitude Pixelization~\cite{Gorski_2005}
(\texttt{HEALPix}\footnote{\url{http://healpix.sourceforge.net}}).

\subsection{Map-making procedure: ISWRS and CMB-lensing}
\label{chap:melita_map}
The procedure followed to build the ISWRS and CMBL potential maps is an adaptation of the one developed in~\cite{Carbone_2008}. To make the ISWRS maps (left column of Figure~\ref{fig:maps}), 
CMB photons are ray-traced along the undeflected line of sight through the 3D field $\dot\Phi$. A similar ray-tracing has been applied to the 3D $\Phi$-grids, in order to produce the CMBL potential maps (right column of Figure~\ref{fig:maps}) from the same realisation of the Universe adopted for the ISWRS maps. To this aim (i.e. producing exactly the same realization of the Universe), the simulated $\dot\Phi$ and $\Phi$-grids have been stacked around the observer, located at $z = 0$, applying the replication and randomisation of the DEMNUni comoving outputs along the line of sight, while keeping periodic boundary conditions in the transverse direction, following~\cite{Carbone_2008}. This particular 3D tessellation scheme is required to avoid both the repetition of the same structures along the line of sight and the generation of artifacts, like ripples, in the simulated deflection-angle field. The latter can be avoided only if the peculiar gravitational potential is continuous transversely to each line of sight. Comparing the left and right columns of Figure~\ref{fig:maps}, it is possible to appreciate at a glance how the ISWRS and the CMBL maps are highly correlated. Comparing the scales of each rows, the slight difference, due to the different $M_{\nu}$ values, emerges.

\subsubsection{Simulated ISWRS and CMB-lensing auto-spectra }
\begin{figure*}[!ht]
    \centering
    \begin{tabular}{cc}
        \includegraphics[width=0.48\textwidth]{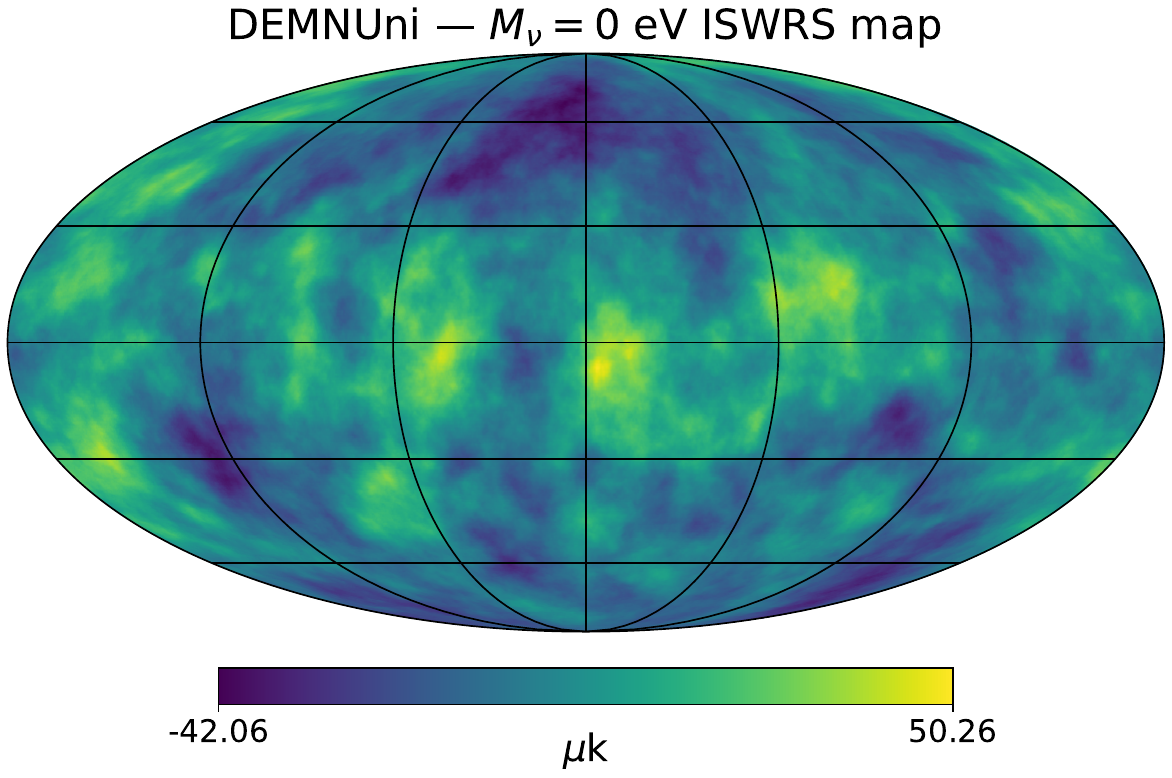} &  \includegraphics[width=0.48\textwidth]{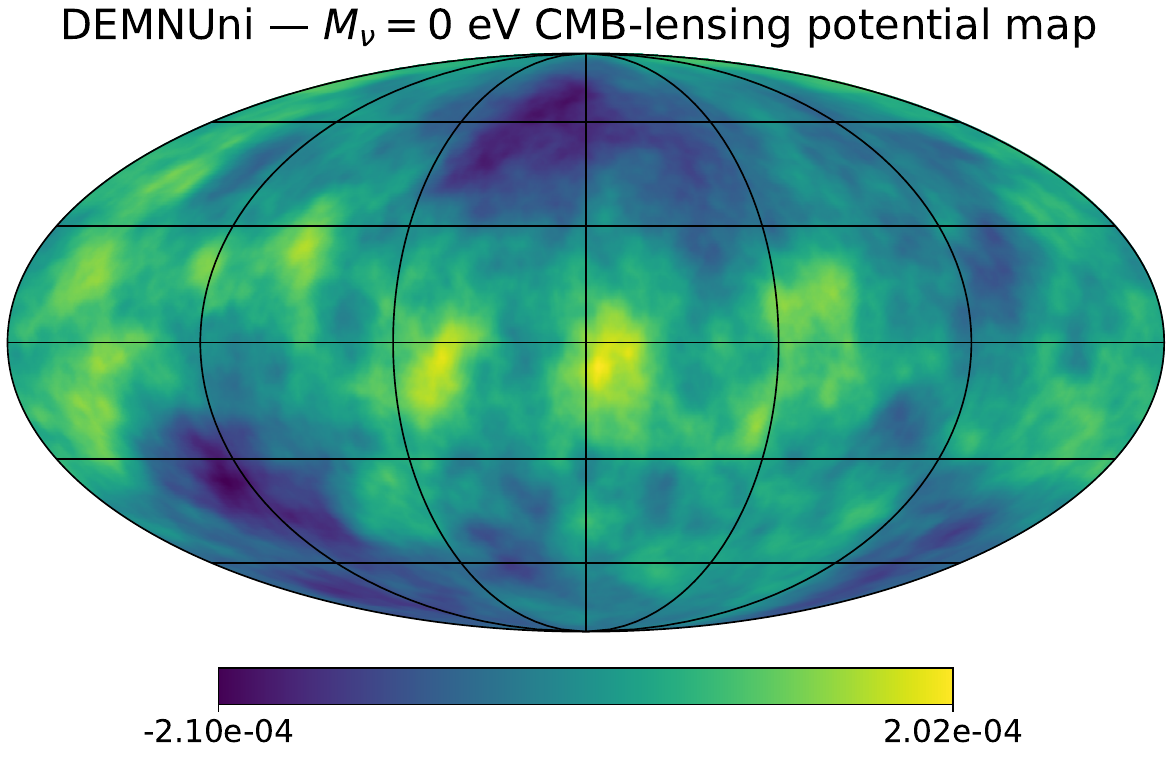}\\
        \includegraphics[width=0.48\textwidth]{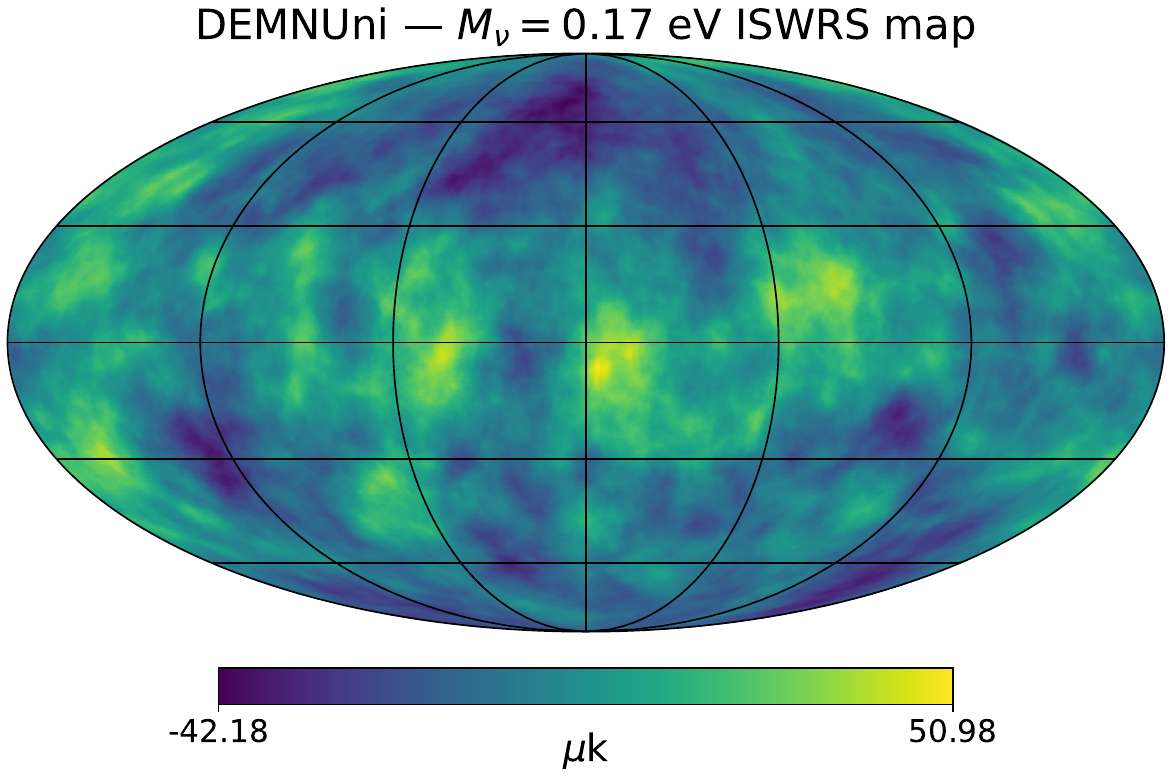} & \includegraphics[width=0.48\textwidth]{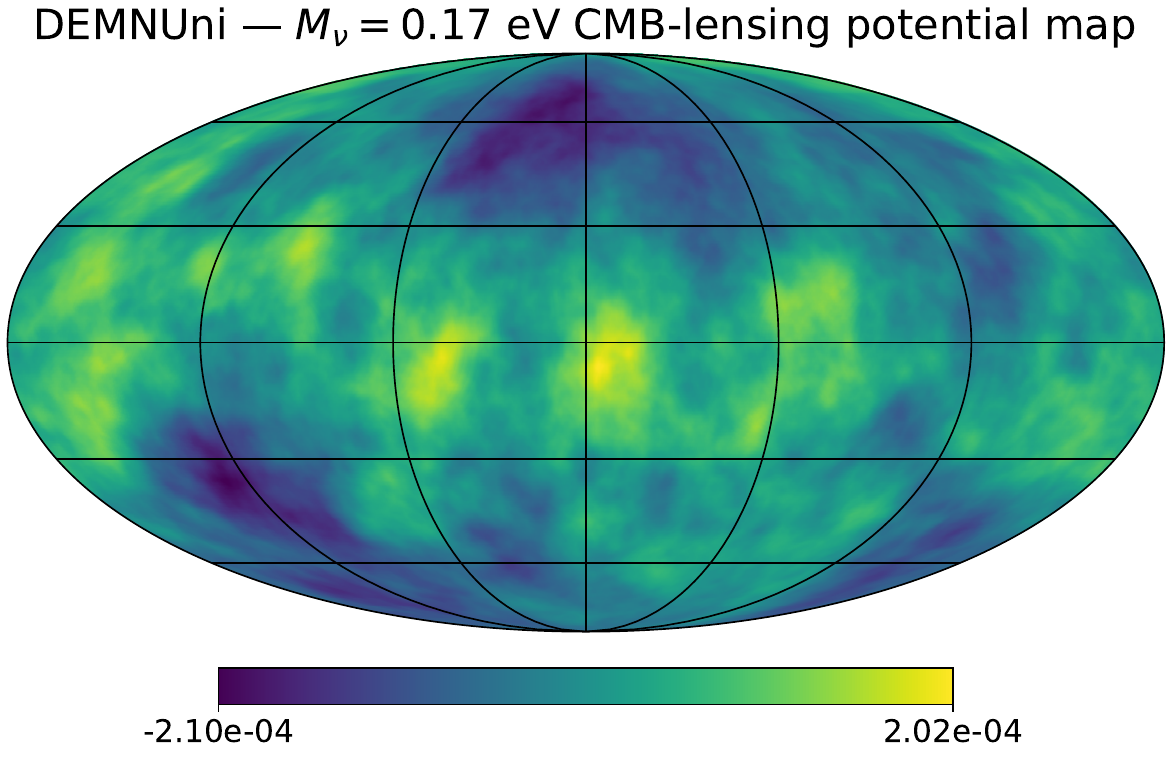} \\
         \includegraphics[width=0.48\textwidth]{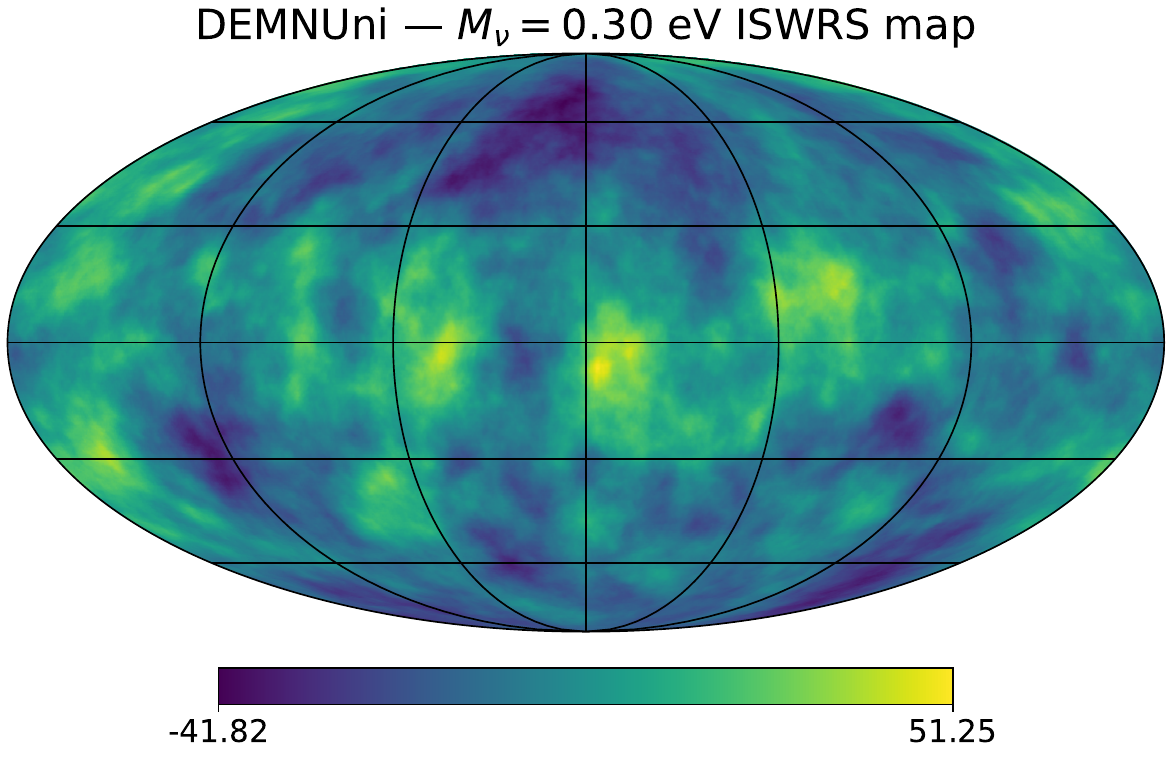} & \includegraphics[width=0.48\textwidth]{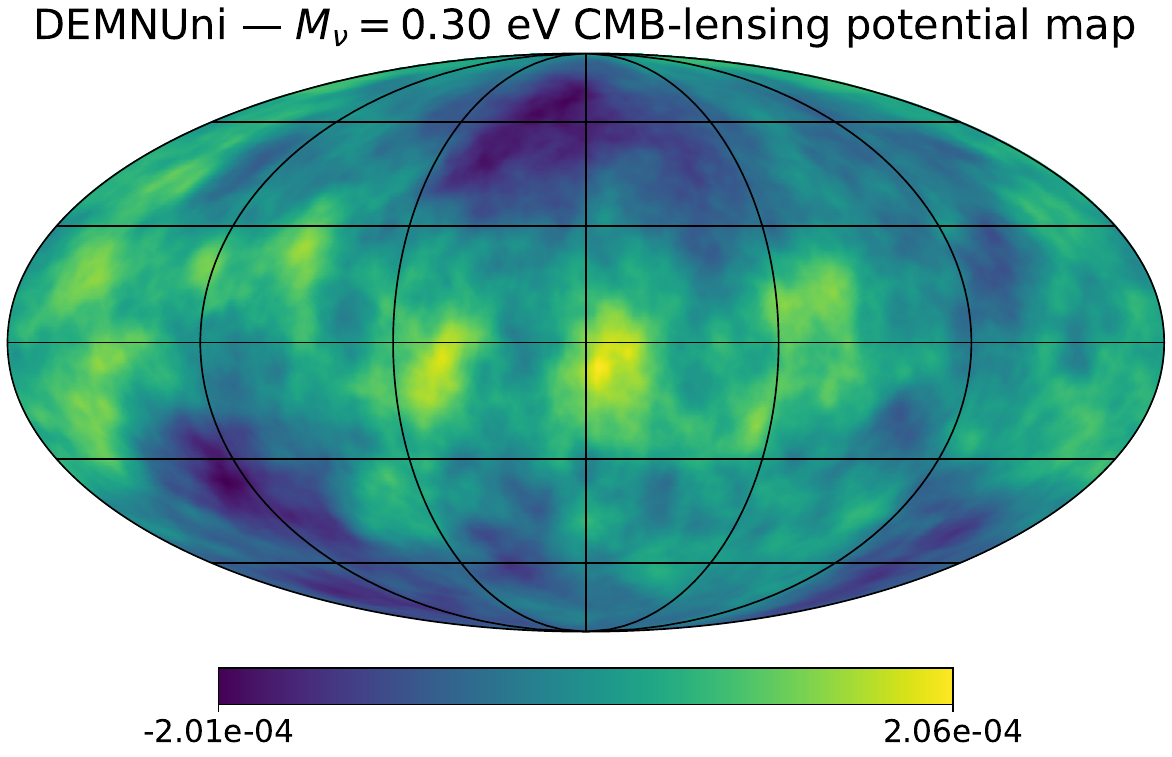} \\
    \includegraphics[width=0.48\textwidth]{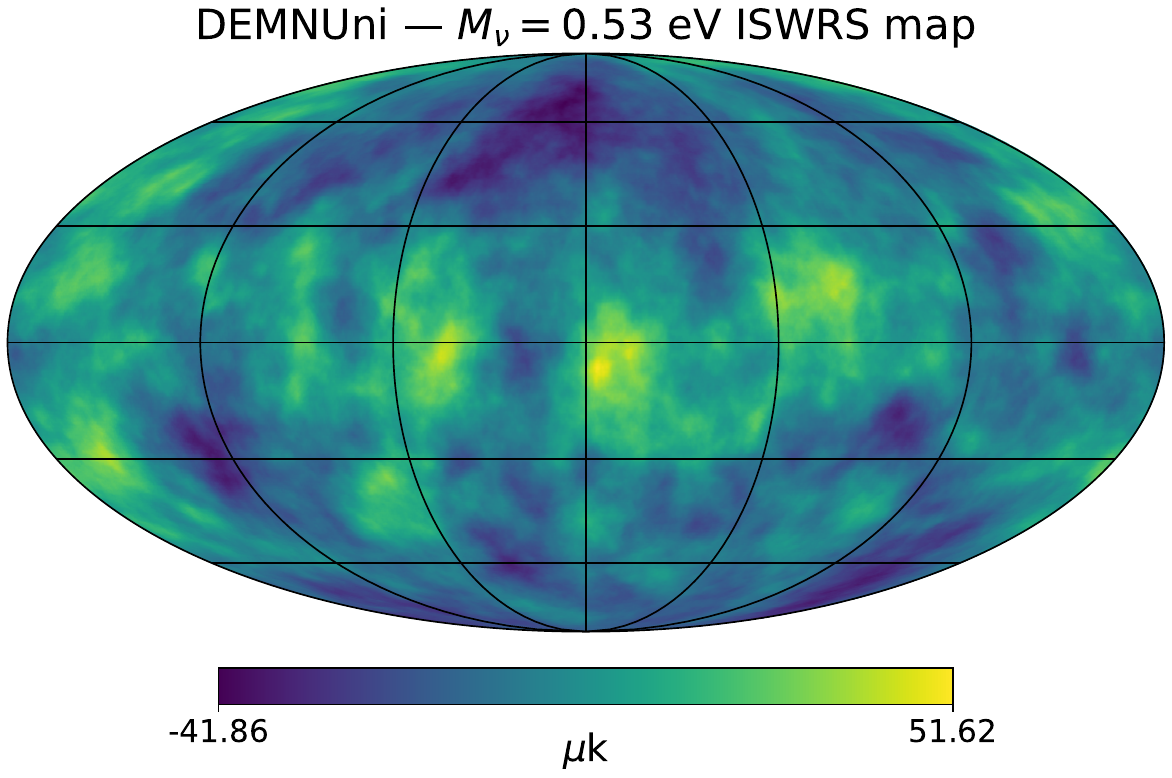} & \includegraphics[width=0.48\textwidth]{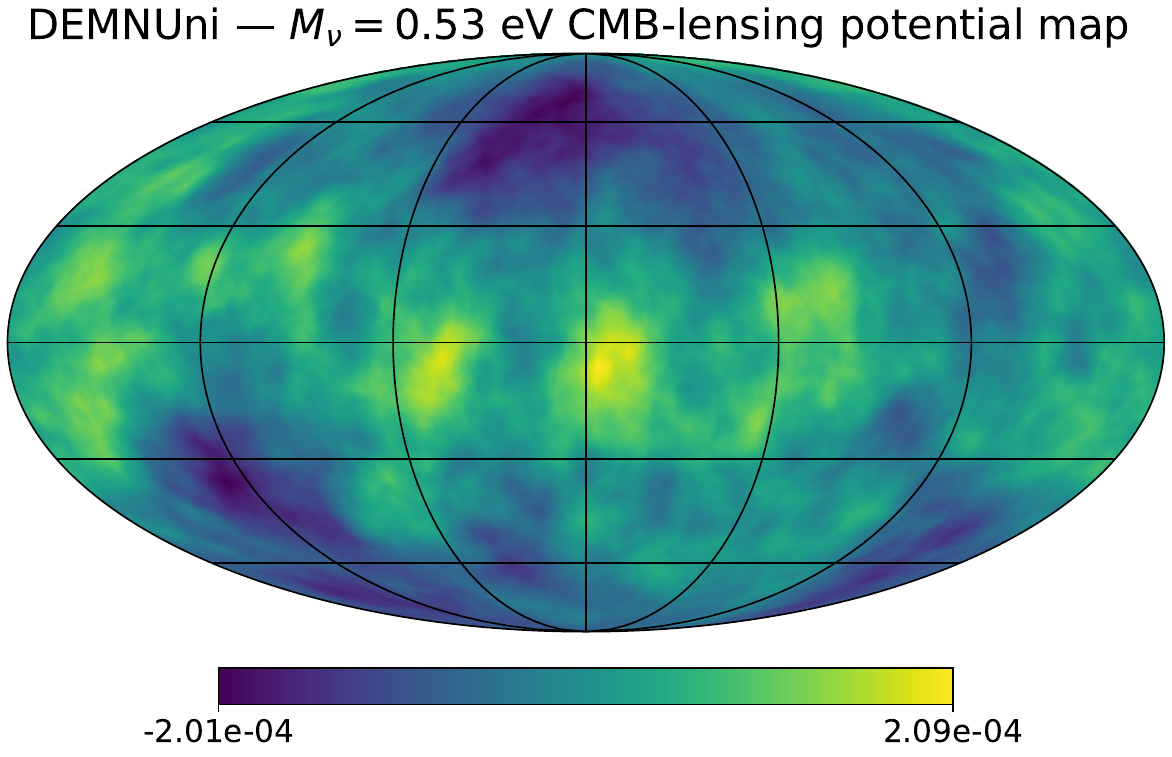}
    \end{tabular}
    \caption{Full-sky maps of ISWRS (\textit{left column}) and CMBL potential (\textit{right column}) obtained from the DEMNUni simulations for the four cosmologies used in this work: $\Lambda$CDM and $\nu\Lambda$CDM with $M_{\nu}=0.17,0.30,0.53\text{ eV}$ (from top to bottom). These maps have a pixel number and resolution given by $N_{\rm side}=2048$ and a RING ordering, following the procedure described in Section~\ref{chap:melita_map}.}
\label{fig:maps}
\end{figure*}

\begin{figure*}[!ht]
\centering
\begin{tabular}{cc}
   \includegraphics[width=0.45\textwidth]{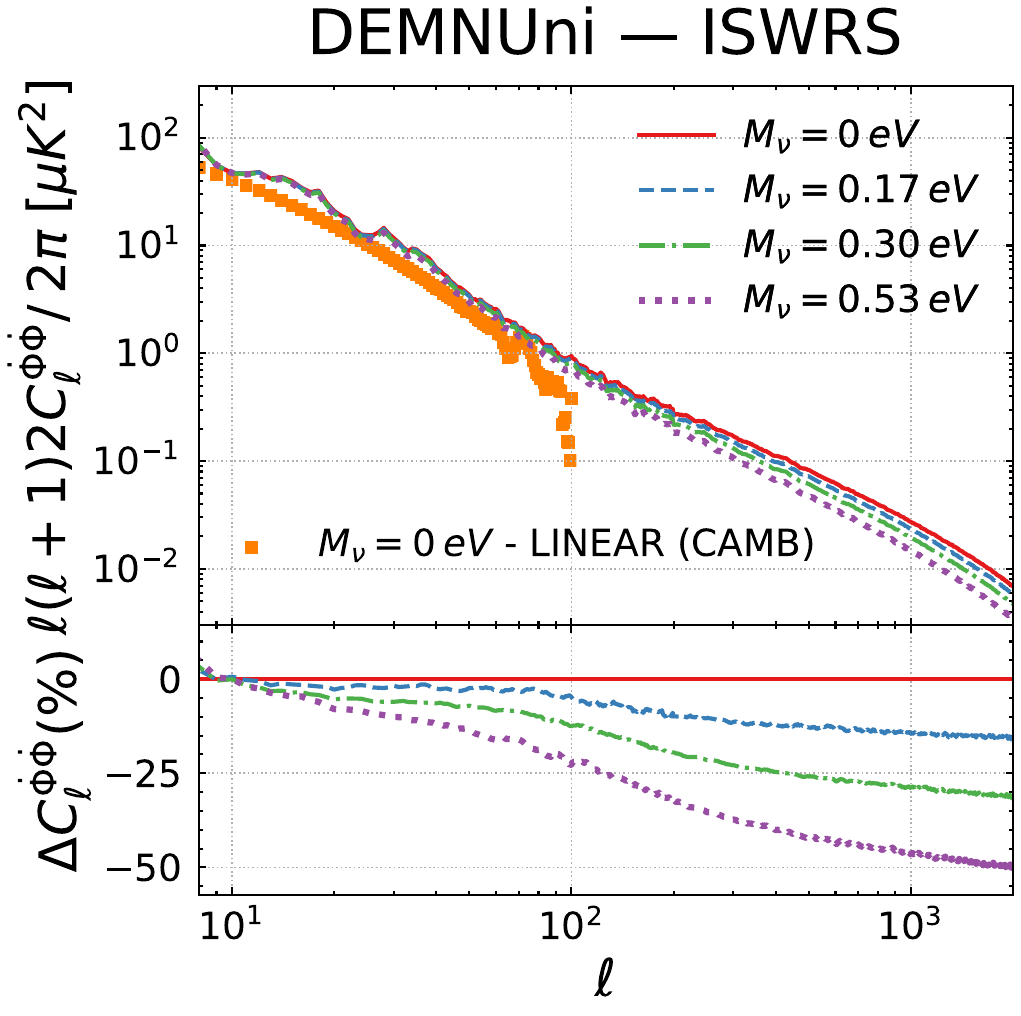}  &  \includegraphics[width= 0.45\textwidth]{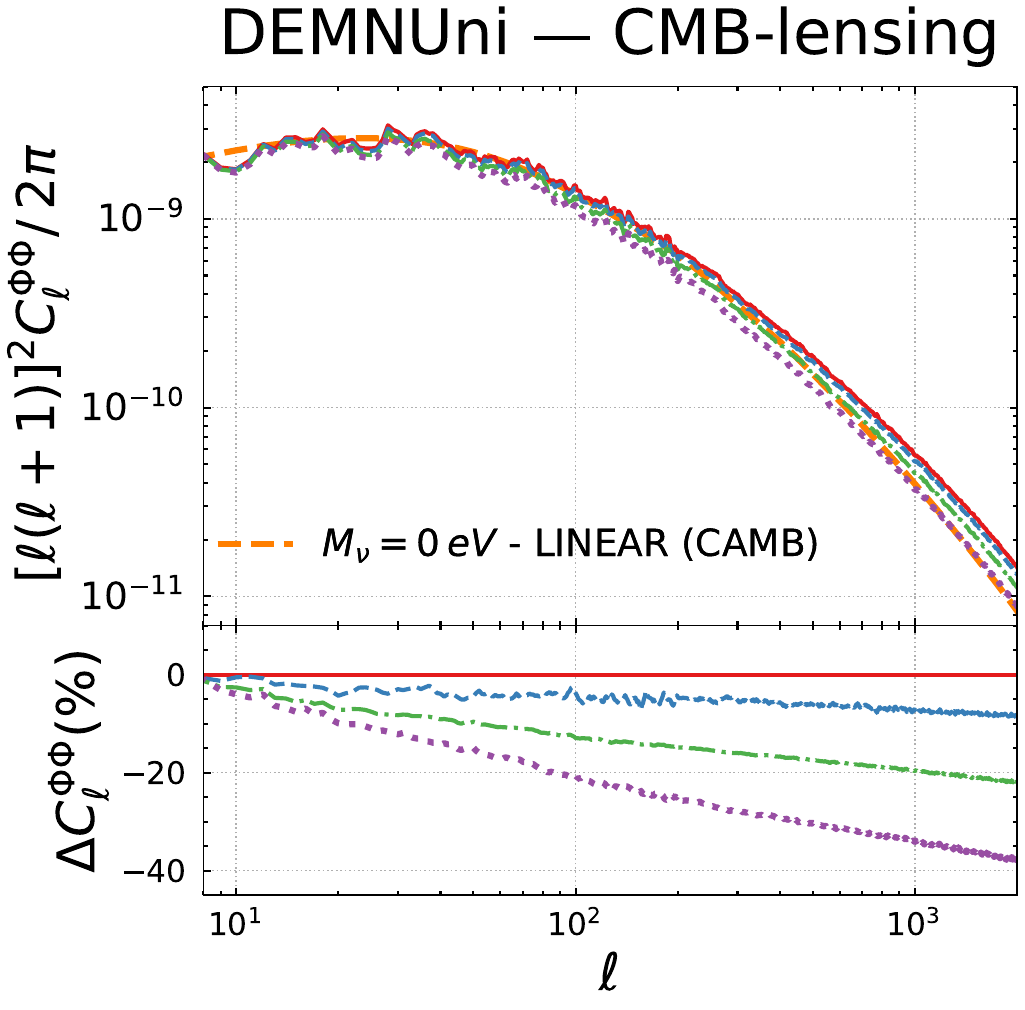} 
\end{tabular}
\caption{\textit{Left}: Angular power spectra of the total ISWRS induced temperature anisotropies computed from DEMNUni maps with $M_{\nu}=0,0.17,0.30,0.53\text{ eV}$ (solid red, dashed blue, dot-dashed green, dotted purple lines, respectively) compared with linear prediction (orange squares). \textit{Right}: Same for angular power spectra of the CMBL potential.
In both subpanels, the percentage relative differences with respect to the massless neutrino case is reported.}
\label{maps1_isw}
\end{figure*}

The power spectra extracted from the maps of  Figure~\ref{fig:maps} are reported in Figure~\ref{maps1_isw}.  
For comparison, the orange squares and the orange dashed lines represent theoretical predictions in the linear regime obtained with \texttt{CAMB} for the massless neutrino case, in order to highlight the differences from the nonlinear modelling.
The subpanels represent the percentage relative differences of the $\nu\Lambda$CDM models with respect to the massless one. There is a good agreement between the simulated ISWRS signals and the ones predicted with \texttt{CAMB} for $\ell <100$ (which roughly corresponds to the transition between the linear and the nonlinear regimes~\cite{Schafer_2006,Schafer_2008}).
Past the linear limit, \texttt{CAMB} predictions fail, given that the RS contribution is not  implemented in the code.
The CMBL potential linear prediction for the massless neutrino case, similarly, perfectly agrees with DEMNUni mocks up to the linear limit and then shows a lack of power with respect to the nonlinear ones. 
\clearpage
Focusing on spectra extracted from maps, it is easy to appreciate at small scales the differences produced in the ISWRS induced temperature anisotropies by free-streaming neutrinos with different total masses. 

Similar effects are noticeable also in the CMBL potential power spectra, considering the percentage relative differences with respect to the massless neutrino case (see subpanels of Figure~\ref{maps1_isw}). 
In particular, as expected, the effect of massive neutrinos produces a suppression of power in the CMBL potential which increases with the neutrino mass and decreases
with the angular scale (at low $\ell$). This effect is reasonably due to the nonlinear excess of suppression of the total matter power spectra with respect to linear expectations in the presence of massive neutrinos, that counteract the nonlinear evolution~\cite{Carbone_2016}. 

\subsection{Map-making procedure: projected galaxy mocks}
\label{chap:galaxy_map}
The procedure followed to construct the galaxy maps is an adaptation of the one developed in~\cite{Calabrese_2015,Fabbian_2018} for convergence maps. 
The convergence maps are extracted with a postprocessing procedure acting on the N-body particle snapshots to create matter particle full-sky lightcones~\cite{Fosalba_2008,Hilbert_2020}. The standard way to build lightcones is to pile up high-resolution comoving snapshots within 
concentric cells to fill the lightcone to the maximum desired source redshift.  With the observer placed at $z=0$, the volume of the lightcone is sliced into full-sky spherical shells of the desired thickness in redshift. All the particles distributed within each of these 3D matter shells are
\begin{figure*}
\centering
\begin{tabular}{cc}
        \includegraphics[width=0.45\textwidth]{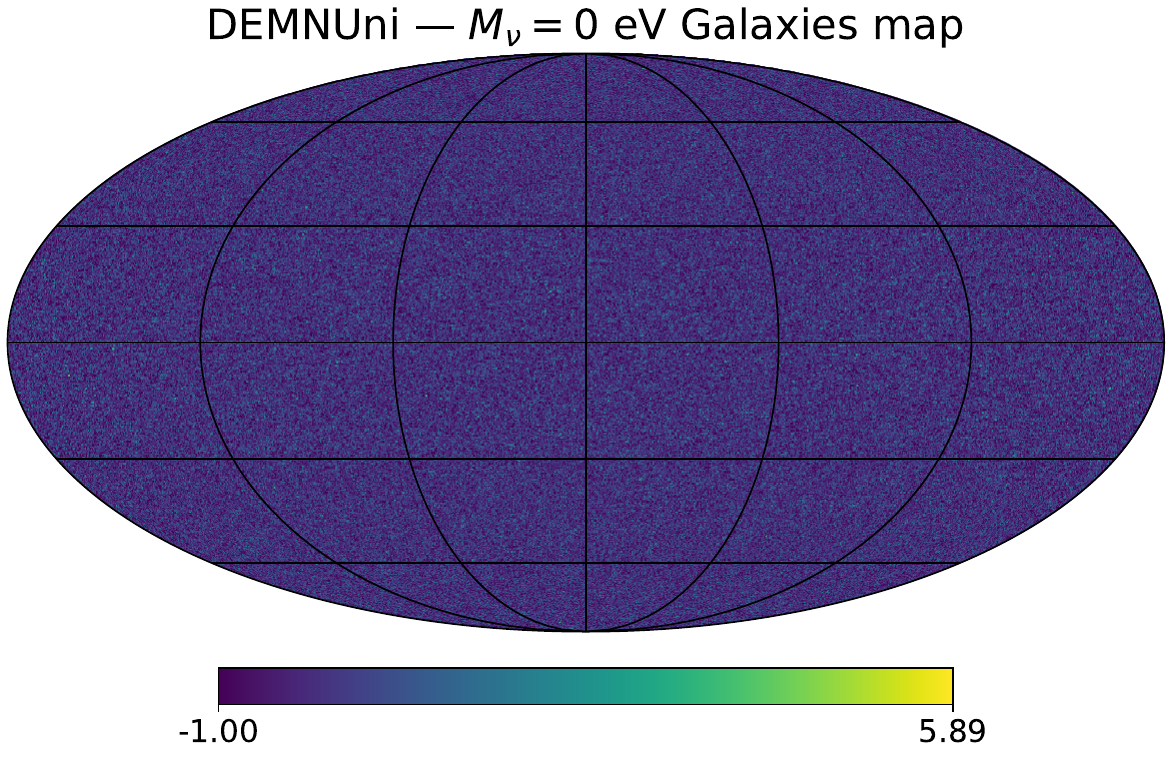} & \includegraphics[width=0.45\textwidth]{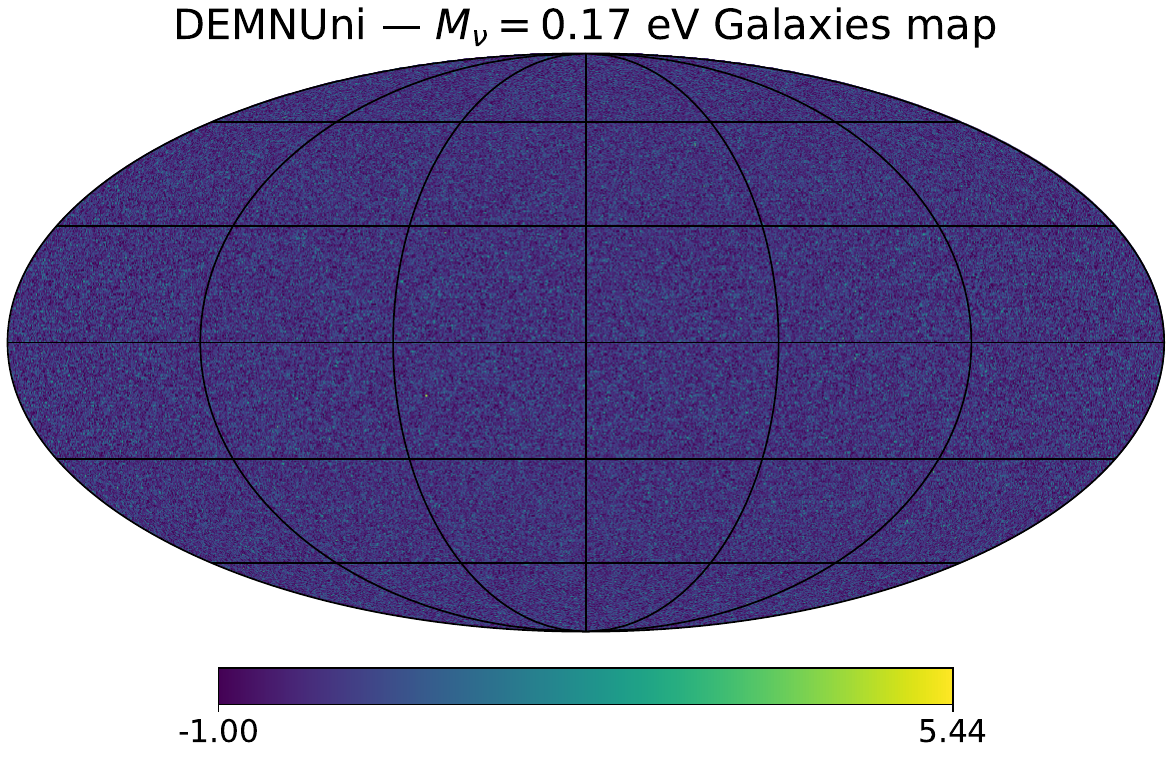}\\
        \includegraphics[width=0.45\textwidth]{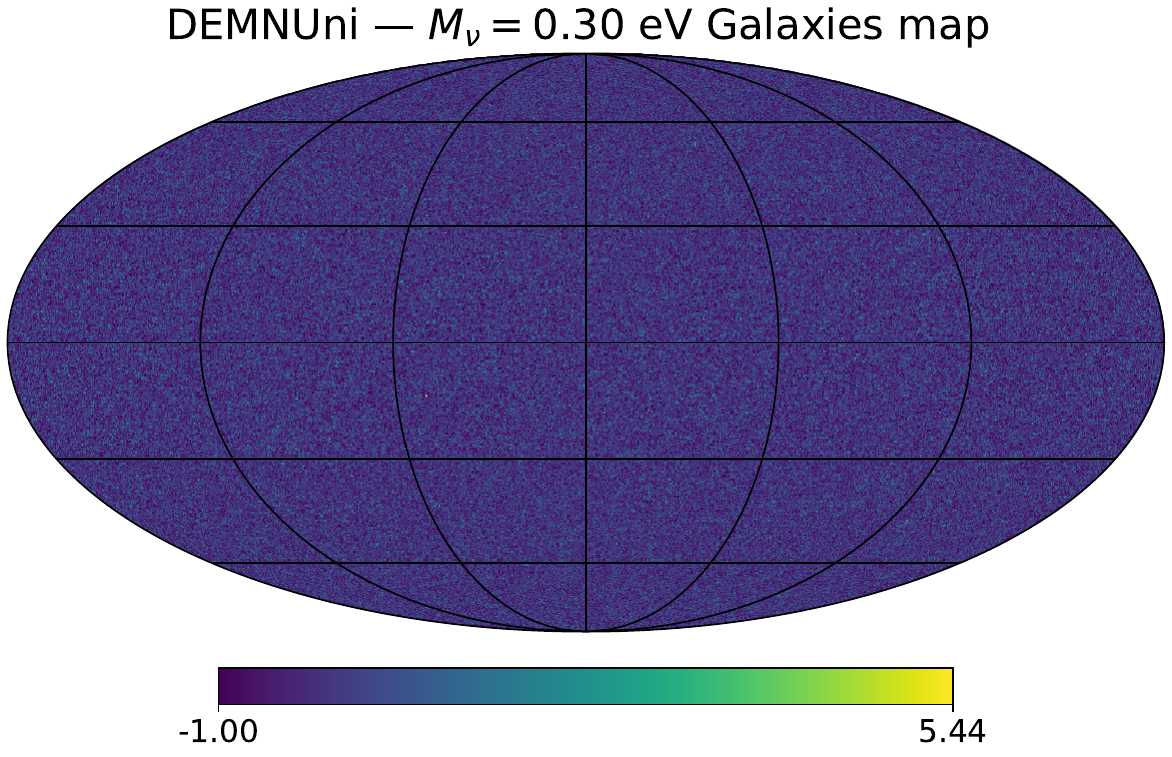} & \includegraphics[width=0.45\textwidth]{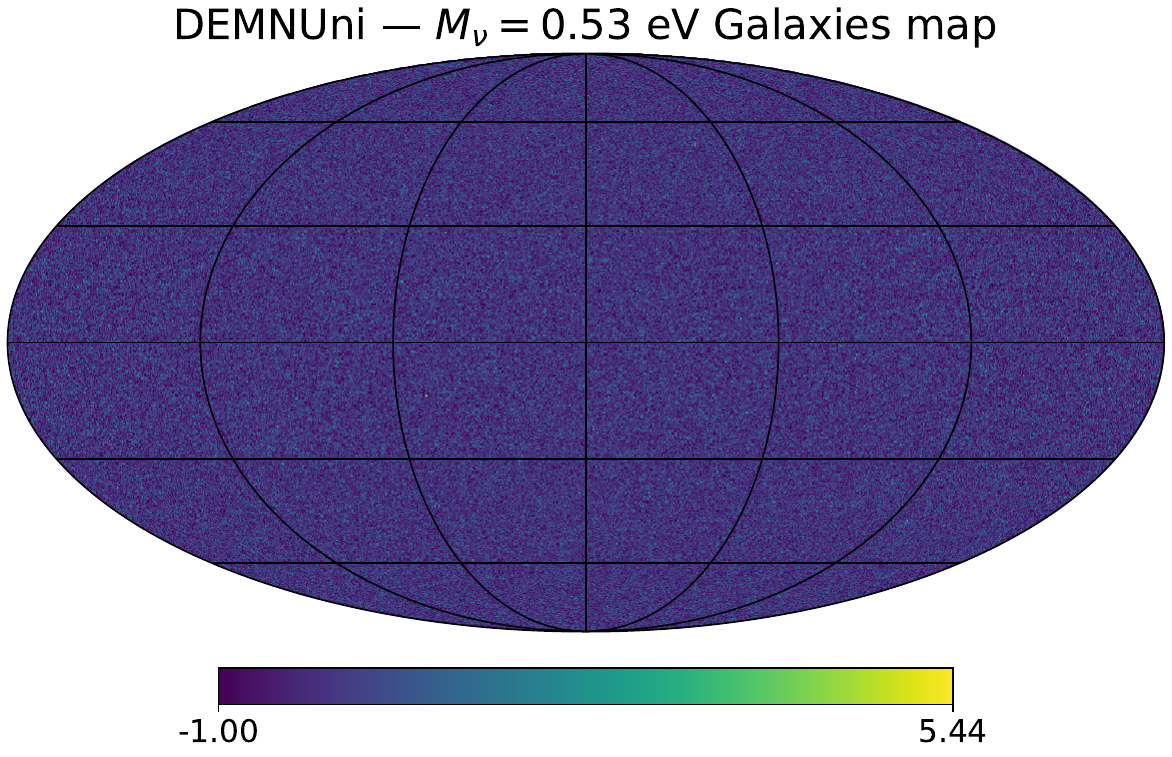}
\end{tabular}
\caption{Full-sky galaxies maps obtained for the four cosmologies used in this work: $\Lambda$CDM and $\nu\Lambda$CDM with $M_{\nu}=0.17,0.30,0.53\text{ eV}$. These maps have a pixel number and resolution given by $N_{\rm side}=2048$ and a RING ordering, following the procedure described in Section~\ref{chap:galaxy_map}.}
\label{fig:galaxy_maps}
\end{figure*}
then projected onto 2D spherical maps, assigning a specific sky pixel to each particle, via \texttt{HEALPix} pixelisation procedure. Then integration along the undeflected line-of-sight is applied to obtain the convergence maps accounting for the lensing efficiency in the desired redshift range. A very similar method has been applied to the DEMNUni galaxies to construct projected galaxies mocks~\cite{Parimbelli_2022} (see Figure~\ref{fig:galaxy_maps}). 
Specifically, in order to reproduce the galaxy distribution in the DEMNUni simulations, the SubHalo Abundance Matching (SHAM) technique~\cite{Carella_in_prep} has been used. The SHAM method assumes a one-to-one relation between a physical property of a dark matter halo/subhalo and an observational property of the galaxy that it hosts. In particular, it assumes that the most massive galaxies form in the dark matter haloes characterised by the deepest potential wells. Galaxies are linked to the corresponding dark matter structures using stellar mass or luminosity as galaxy property and a measure of halo mass or the circular velocity, as halo property (i.e. a proxy of the depth of the local potential well). Here we link the stellar mass to the halo mass using the Moster et al. (2010)~\cite{Moster_2010} parameterised stellar-to-halo mass relation, fitted against the DEMNUni simulations. As the SHAM galaxy catalogues are fitted against observations, they guarantee an high-degree of agreement with real data from galaxy surveys~\cite{Shankar_2006, Girelli_2020}. 

We do not expect the particular galaxy modelling, and in particular here the SHAM parameters, to significantly affect the sign inversion position or to improve the sensitivity to $M_{\nu}$. In fact, we will show in the following Sections that the cross-correlation between ISWRS and CMBL and the one between ISWRS and galaxy clustering share a very similar trend as far as the $l_{inv}$ is concerned, even if the CMBL does not involve any galaxy bias. Moreover, the SHAM parameters may at most affect the galaxy bias function and its scale-dependence, and we will show in Section~\ref{sec:XC-TG} that the sign inversion position is not very sensitive to the particular bias modelling but rather to the bias amplitude, i.e. the galaxy population considered and its evolution with $z$.

Applying to the SHAM galaxy comoving snapshots the same stacking technique as for particle comoving snapshots, we have generated full-sky mock galaxy catalogues from the DEMNUni simulations in different cosmological models. 

\subsubsection{Galaxy bias and selection function from DEMNUni simulations}
The galaxy bias and selection function necessary to compute the analytical predictions of the cross-spectra between ISWRS and galaxies (see Equation~\eqref{ctg}) have been directly extracted from the DEMNUni simulations. This guarantees a more accurate comparison between mocks and analytical predictions. 
\paragraph{Galaxy selection function}
The galaxy selection function extracted from the DEMNUni $\Lambda$CDM map can be modelled via the following equation:
\begin{equation}
\label{eq:demnuni_n}
    n_{\rm DEMNUni}(z) = A\left(\frac{z}{z_{0}}\right)^{B} \exp\Big[ -\Big(\frac{z}{z_{0}} \Big)^{C}\Big]\,,
\end{equation}
where A = 1.89, B = 1.95, C = 1.41 and $z_{0} = 0.88$. 

This is the $n(z)$ used to compute all the theoretical predictions to be compared with DEMNUni data, also for massive neutrino cases. This is because galaxy counts vary mostly because of the variation of the simulated cosmological volume, which depends on $H(z)$ via $\Omega_{m}$ - that has been fixed to 0.32 for all the simulations (and consequently for all the theoretical predictions) - and via the DE EoS, whose presence has not been taken into account in this work. As a result, the normalised galaxy counts do not undergo any significant change due to the variation of $M_{\nu}$, and consequently it is possible to exploit the same $n(z)$ to compute the theoretical predictions even in the massive neutrino case. We have verified this by measuring the galaxy counts from the DEMNUni catalogues with different values of $M_\nu$, obtained via the SHAM technique applied to COSMOS2020 and SDSS observations ~\cite{Carella_in_prep}.
\begin{figure*}[!ht]
\centering
\begin{tabular}{cc}
    \includegraphics[width=0.45\textwidth]{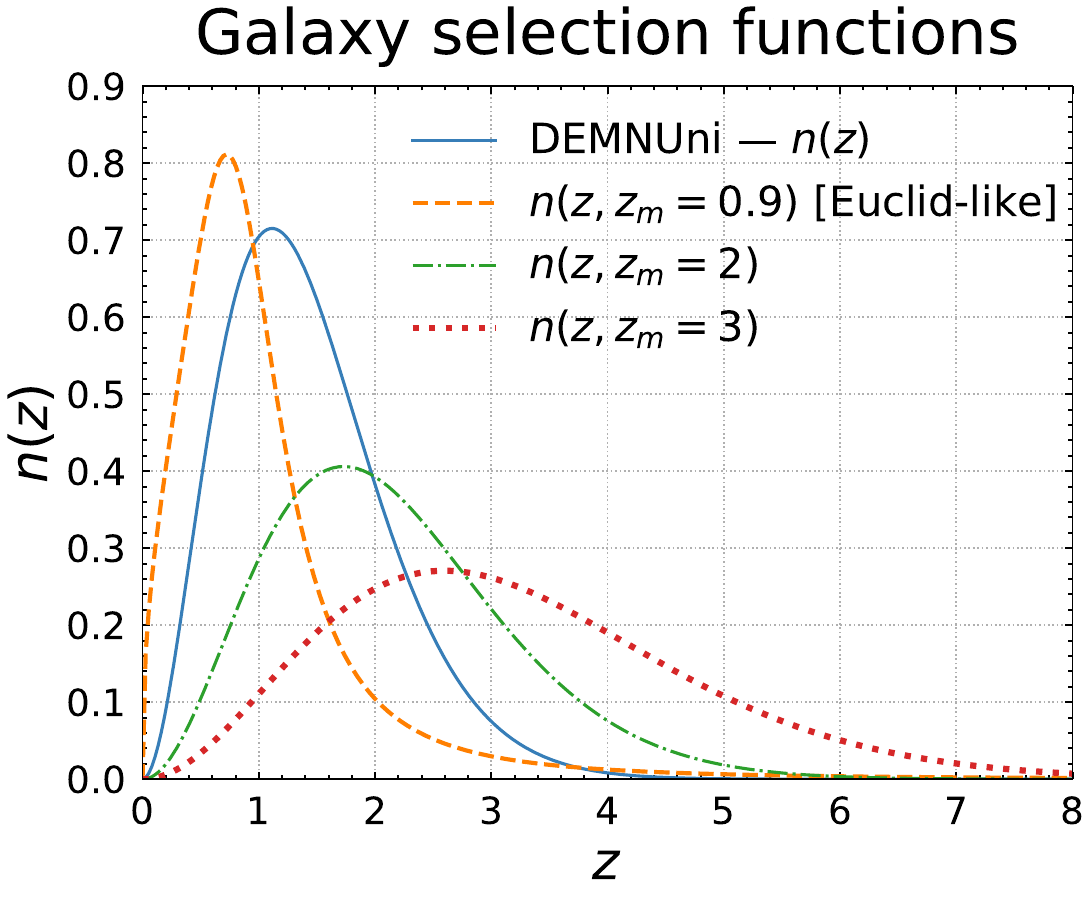} &
    \includegraphics[width=0.45\textwidth]{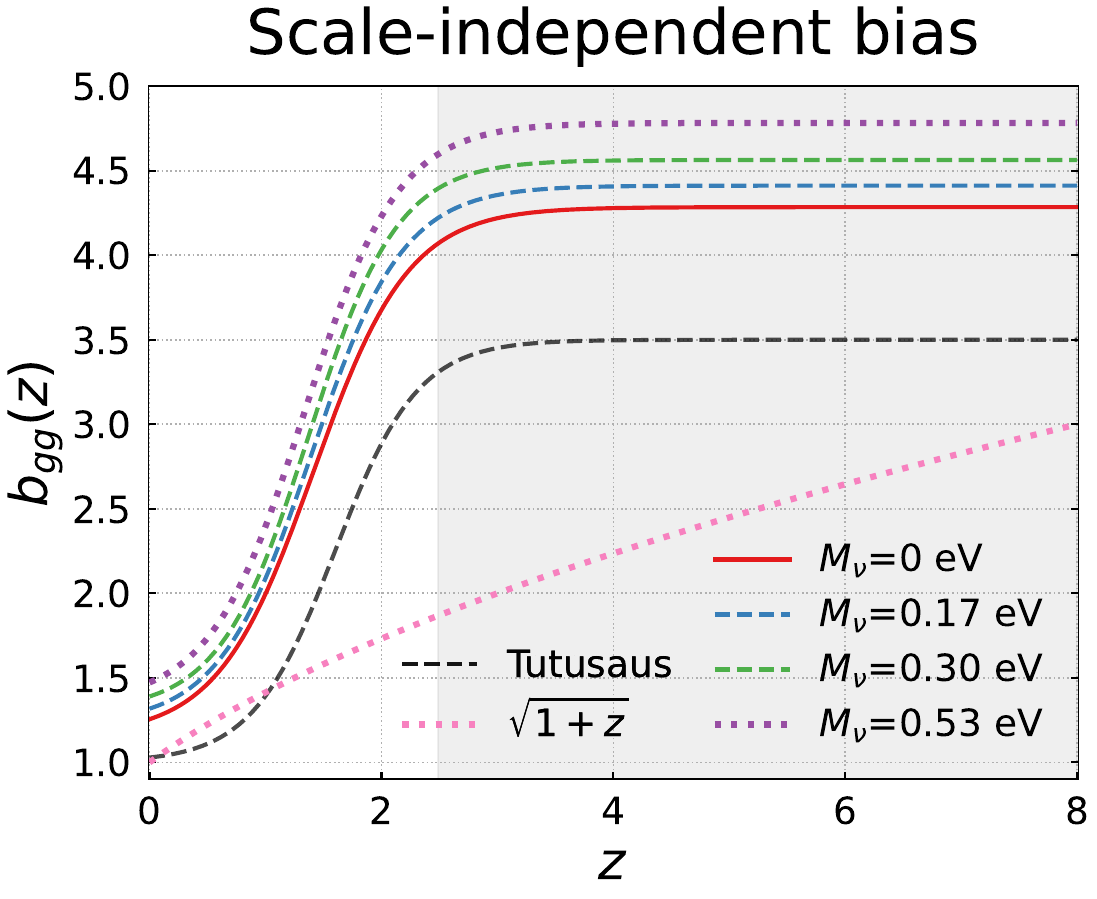}
\end{tabular}
\caption{ \textit{Left}: Galaxy selection functions: the solid blue line is extracted from DEMNUni data; the dashed orange line is $n(z, z_{m})$ from Equation~\eqref{eq:Euclid_n} with $z_{m} = 0.9$, which is the one assumed for the photometric \textit{Euclid} survey; the dot-dashed green and the dotted red lines are $n(z, z_{m})$ from Equation~\eqref{eq:Euclid_n} with $z_{m} = 2, 3$, respectively~\cite{Lesgourgues_2008}. \textit{Right}: Scale-independent galaxy bias functions: 
the dashed black line is the Tutusaus bias~\cite{Tutusaus_2020}; the solid red, dashed blue, dot-dashed green and dotted purple lines are $b_{gg}(z, M_\nu)$ in Equation~\eqref{eq:euclid_bias} for $M_{\nu}=0,0.17,0.30, 0.53$ eV, respectively; finally the dotted pink line is $b_{gg}(z)=\sqrt{1+z}$ used in~\cite{EuclidVII_2020}. The shaded grey area highlights the extrapolation of the galaxy bias function  above the redshifts tested against the \textit{Euclid} Flagship 1  simulation~\cite{Tutusaus_2020}.}
\label{n_b}
\end{figure*} 

In Figure~\ref{n_b}, together with the DEMNUni $n(z)$, we show the galaxy selection function adopted in \cite{Lesgourgues_2008}:
\begin{equation}
\label{eq:Euclid_n}
    n_{\rm}(z, z_{m}) = \frac{3 \cdot (z/z_{0})^{\alpha}}{2 \cdot z_{0}} \text{exp} \Big[-\Big(\frac{z}{z_{0}}\Big) ^{\beta} \Big] \,,
\end{equation}
where $\alpha$ = 2, $\beta$ = 1.5, $z_{0}$ = $z_{m}/\sqrt{2}$ and $z_{m} = 0.9, 2, 3$. The case $z_{m}=0.9$ corresponds to the galaxy selection function expected for the photometric \textit{Euclid} survey~\cite{Laureijs_2011, EuclidVII_2020}. Equation~\eqref{eq:Euclid_n} (with $z_{m} = 0.9, 2, 3$) has been used in Section~\ref{sec:forecasts} to produce forecasts for $z \ge 2$, i.e. for redshifts larger than the ones investigated in the DEMNUni maps. 

\begin{figure*}[h!]
\centering
\begin{tabular}{cc}
   \includegraphics[width=0.48\textwidth]{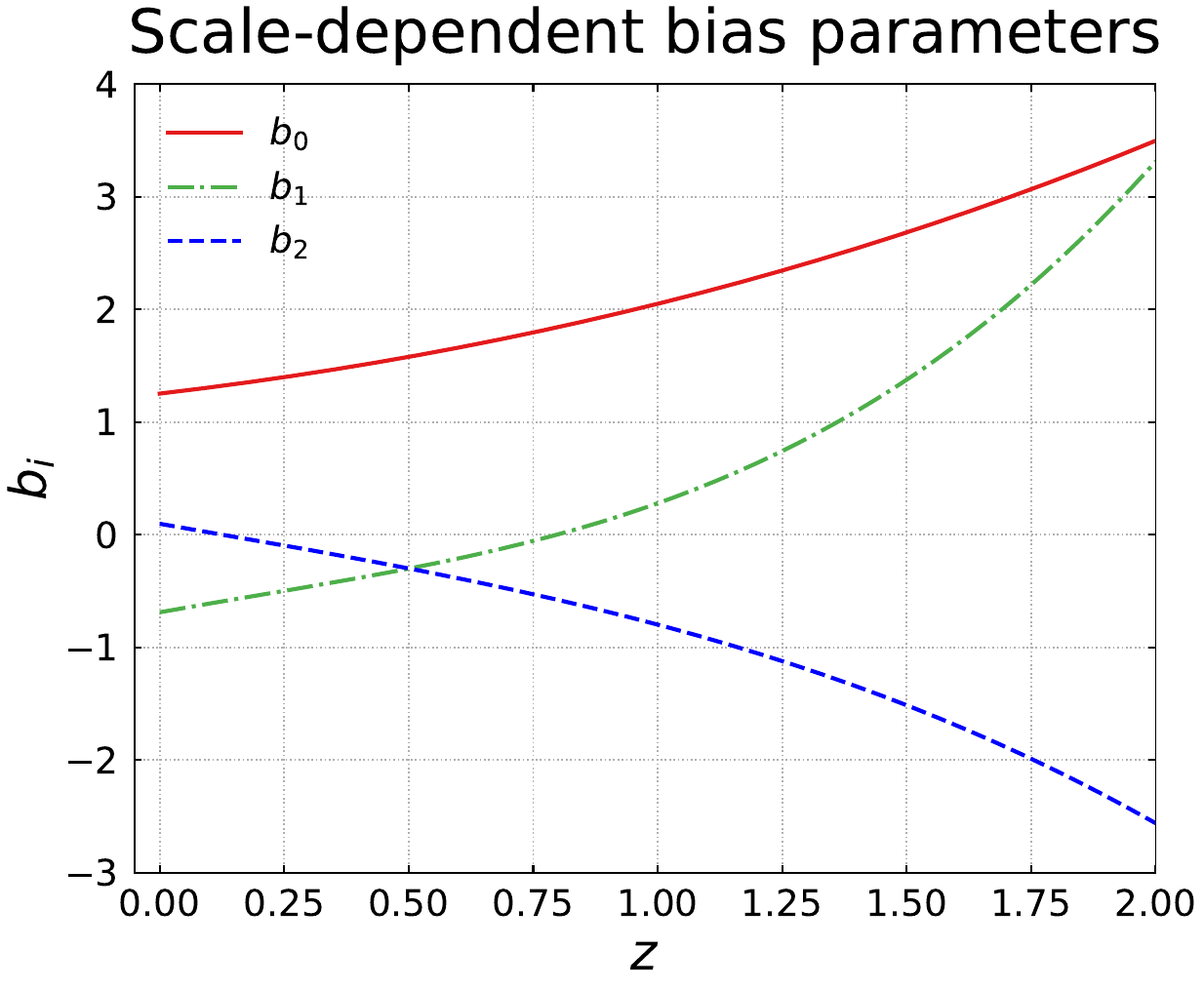} &
    \includegraphics[width=0.48\textwidth]{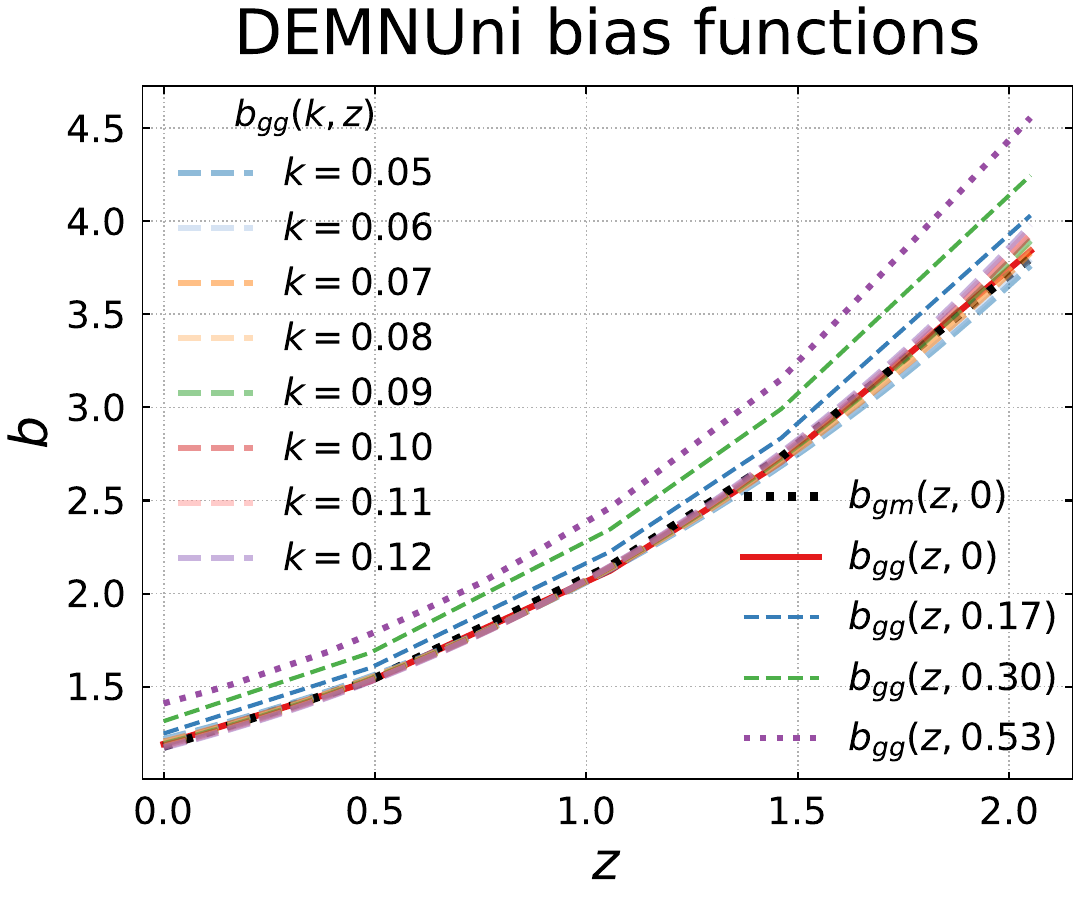}
\end{tabular}
\caption{ \textit{Left}: Evolution of bi(z) parameters from Equation  of $b_{i}(z)$ parameters from Equation~\eqref{eq:bnl}, in the redshift range $0.02\le z \le1.89$ of the DEMNUni maps.
\textit{Right}: Linear interpolation of $b_{gg}(z, M_{\nu})$ and $b_{gm}(z, 0)$ from Equations~\eqref{eq:bgg}-\eqref{eq:bgm}, respectively, and $b_{gg}(k,z)$ from Equations~\eqref{eq:bnl}, computed for large scales in the range $0.05 < k < 0.12 \text{ Mpc}^{-1}$ .}
\label{bnl}
\end{figure*}

\paragraph{Scale-independent bias}
The galaxy scale-independent bias has been computed, for each $M_\nu$ value, as the linear interpolation in $0<z<2$ of the bias measurements obtained as the square root of the ratio between the galaxy auto power spectrum, $P_{gg}(k,z)$, and the matter power spectrum, $P_{mm}(k,z)$, from the DEMNUni galaxy and dark matter comoving snaphots, respectively:
\begin{equation}
\label{eq:bgg}
    b_{gg} = \Bigg\langle\sqrt{\frac{P_{gg}}{P_{mm}}}  \Bigg\rangle_{k} \,.
\end{equation}
Here the average has been computed on large linear scales, specifically in the range $0.05<k<0.12$ $\text{Mpc}^{-1}$. The linear interpolation in $z$ of Equation~\eqref{eq:bgg} for $M_{\nu}=0, 0.17, 0.30, 0.53$ eV, that from now on we call $ b_{gg}(z,M_\nu)$, are reported in the left panel of Figure~\ref{bnl}.
We use the measurements in Equation~\eqref{eq:bgg} to fit, for the different values of $M_{\nu}$, the coefficients in the following functional form adopted to produce the forecasts in Section~\ref{sec:forecasts} at $z>2$:
\begin{equation}
\label{eq:euclid_bias}
    b^{Tut}_{gg}(z,M_\nu) = A + \frac{B}{1 + \exp[-(z-D){C}]} \,,
\end{equation}
where the coefficients A, B, C and D depend on the neutrino mass, and the specific values A = 1.0, B = 2.5, C = 2.8, D = 1.6 correspond to the galaxy bias model from the \textit{Euclid} Flagship 1 simulation, reported in~\cite{Tutusaus_2020}, and shown in the right panel of Figure~\ref{n_b}. We call Equation~\eqref{eq:euclid_bias}, for the different values of $M_\nu$, the ``a la Tutusaus'' bias.

When we compare our theoretical predictions with the DEMNUni data, we consider also, for the massless neutrino case alone, the galaxy scale-independent bias computed as the linear interpolation in $0<z<2$ of the bias measurements obtained as the ratio between the galaxy-matter cross-spectrum, $P_{gm}(k,z)$, and the matter power spectrum, $P_{mm}(k,z)$, from the DEMNUni galaxy and dark matter comoving snaphots:
\begin{equation}
\label{eq:bgm}
    b_{gm} = \Bigg\langle\frac{P_{gm}}{P_{mm}}  \Bigg\rangle_{k} \,.
\end{equation}
Here the average has been computed again over the range $0.05<k<0.12$ $\text{Mpc}^{-1}$ and its linear interpolation is reported in the left panel of Figure~\ref{bnl}. Since it is affected by a lower level of shot noise, $b_{gm}(z, M_\nu = 0 )$ should allow us to describe better the connection between matter and galaxies, and consequently to improve the cross-correlation reconstruction. However, the results obtained do not differ significantly from those obtained using $b_{gg}(z, M_{\nu} = 0)$, as we will show in Section~\ref{test_bias_mnu0}.

\paragraph{Scale-dependent bias}
A scale-dependent bias fit has been obtained from the DEMNUni massless neutrino mocks using the functional form below~\cite{Modi_2020}:
\begin{equation}
    b_{gg}(k,z) = b_{0}(z)+b_{1}(z)k+b_{2}(z)k^2 \,,
    \label{eq:bnl}
\end{equation}
where $b_{0}$ is the scale-independent bias, $b_{2}$ is a correction due to the peak theory and $b_{1}$ does not have a physical meaning, but it is necessary to improve the fit. These parameters have been obtained fitting Equation (3.6) up to $k=1\text{ Mpc}^{-1}$ against the DEMNUni scale-dependent bias measurements.  
In Figure~\ref{bnl}, we show the evolution of the $b_{i}$ parameters in the left panel, while on the right we show $b_{gg}(k,z)$ as a function of $z$ at some chosen wavenumbers in the range of scales used to measure the scale-averaged  $b_{gg}(z, M_\nu=0)$.

\subsubsection{Simulated projected galaxy auto-spectra}
\begin{figure*}[!ht]
 \centering
\includegraphics[width=0.6\textwidth]{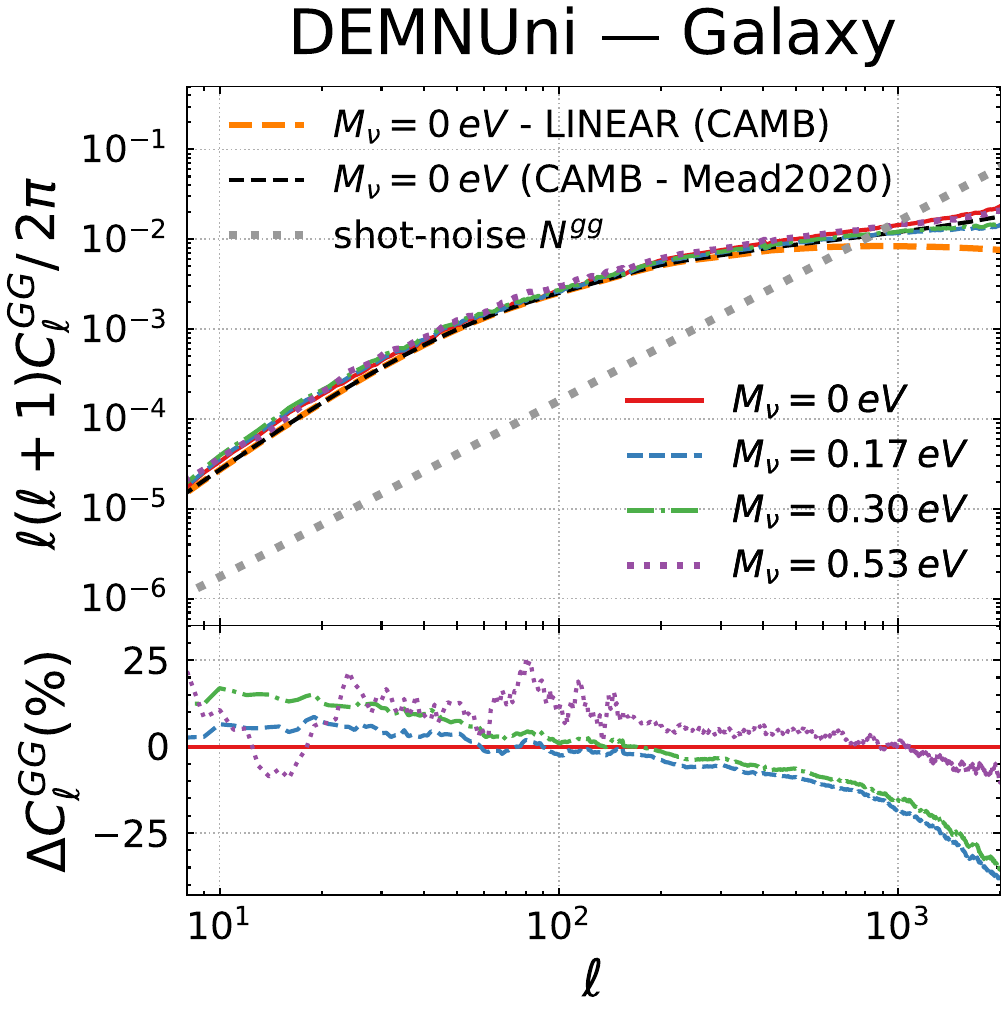}
\caption{Galaxy angular auto-spectra extracted from the DEMNUni maps for $M_{\nu}=0,0.17,0.30,0.53$ eV (solid red, dashed blue, dot-dashed green, dotted purple lines, respectively) with shot-noise (dotted grey line) subtracted, compared with theoretical predictions in the linear (dashed orange line) and \citetalias{Mead_2021} nonlinear (dashed black line) regimes. In the subpanel, the percentage relative difference with respect to the massless neutrino case is reported.}
\label{c_gg}
\end{figure*}

In Figure~\ref{c_gg}, we report the spectra extracted from the galaxy maps of Figure~\ref{fig:galaxy_maps} after subtracting the shot-noise value.

In addition to the DEMNUni maps spectra, we report the theoretical predictions of the $C_{\ell}^{gg}$ computed in both the linear and nonlinear regimes, for the massless neutrino case, via~\cite{Calabrese_in_prep}:
 \begin{equation}
     C_{\ell}^{gg} = \int_{z_{\rm min}}^{z_{\rm max}} {\rm d}z\, \frac{H(z)}{c \chi^{2}(z)} [W^{g}(z)]^{2}P_{\delta\delta}(k,z) \,,
     \label{c_l_gg}
 \end{equation}
 where:
 \begin{equation}
   \label{W_eq}
     W^{g}(z) = n(z) \cdot b(z)
     \,,
 \end{equation}
and for $n(z)$ and $b(z, M_{\nu} = 0)$ we use Equation~\eqref{eq:demnuni_n} and Equation~\eqref{eq:bgg}, respectively.

Both linear and nonlinear predictions do not perfectly agree with simulated spectra on large scales. This is due to the limited dimension of the simulation box, which does not allow to represent those scales. Conversely, on small scales there is a good agreement with the nonlinear prediction and the expected disagreement with the linear one. 

Despite the presence of shot-noise in the DEMNUni maps, its effect is negligible when cross-correlating these maps with the ISWRS maps (since the noises of the two fields do not correlate). Furthermore, this analysis does not aim to reveal a perfect match between the simulations and the analytical predictions also because of inaccuracies in the adopted nonlinear modelling.

The subpanel in Figure~\ref{c_gg} represents the percentage relative differences of the $\nu\Lambda$CDM models with respect to the massless neutrino one. Here the larger $M_{\nu}$ is, the smaller the difference with respect to the massless neutrino case is. This is because the matter perturbation suppression due to more massive neutrinos is compensated by a stronger clustering~\cite{Marulli_2011}.

\section{The ISWRS cross-correlation with the CMB-lensing potential}
\label{sec:XC-TP} 
\begin{figure*}[!ht]
\centering
\includegraphics[width=0.90\textwidth]{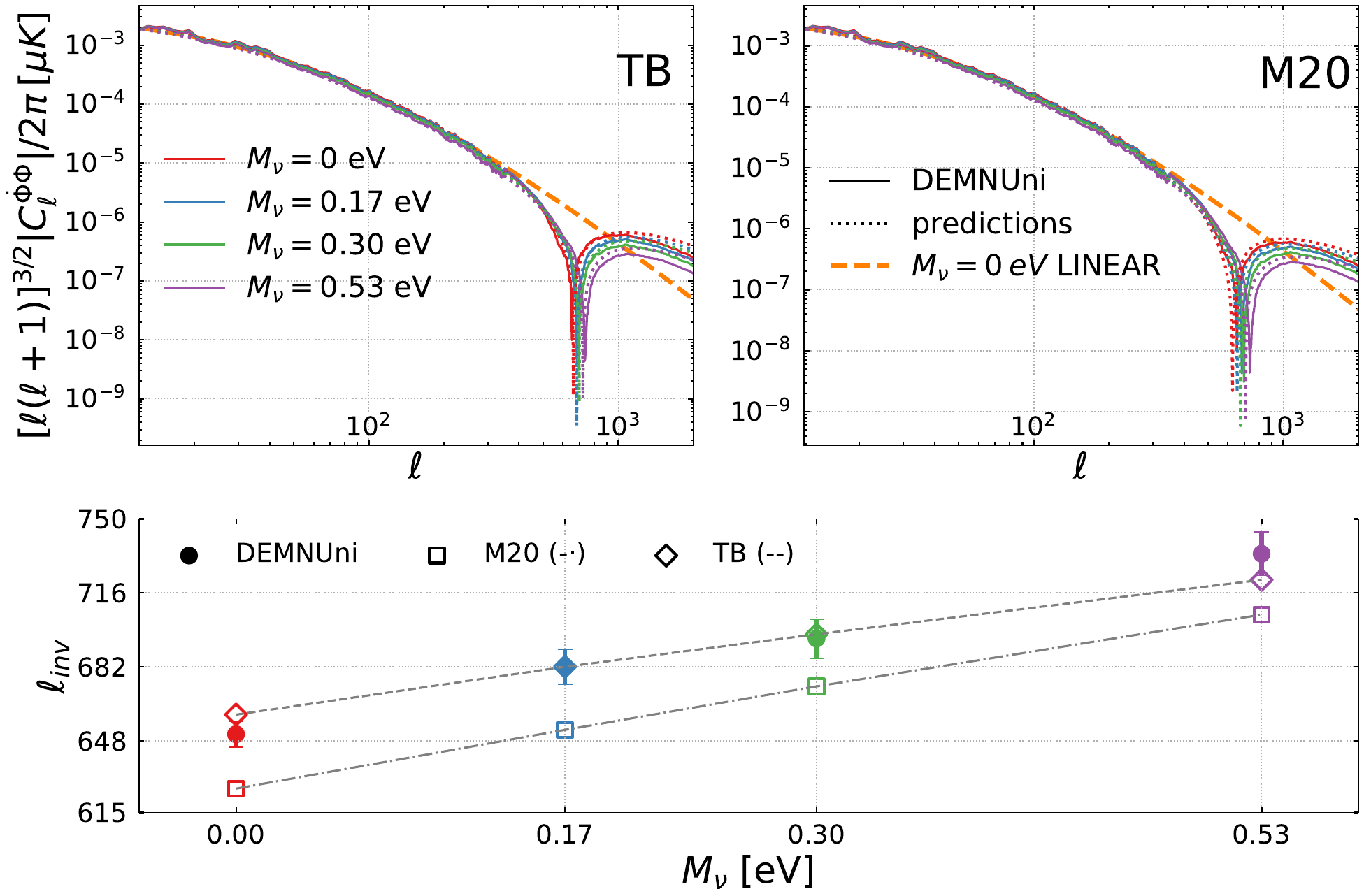}
\caption{ISWRS--CMBL angular cross-spectra  for $M_{\nu}=0,0.17,0.30,0.53\text{ eV}$ (solid red, blue, green and purple lines, respectively) from the DEMNUni maps compared to the theoretical predictions in the linear (dashed orange line) and nonlinear (dotted lines) regimes, obtained using the \citetalias{Takahashi_2012} model (left) and the \citetalias{Mead_2021} model (right), respectively.  The subpanel shows the behaviour of $\ell_{inv}$ as a function of $M_{\nu}$ for the two nonlinear models, compared with the measurements from the DEMNUni simulations.}
\label{cross_TP}
\end{figure*}

The presence of the sign inversion at large multipoles and its shift due to the variation of $M_{\nu}$ in the cross-power spectrum between the ISWRS effect and the CMBL was already verified in~\cite{Carbone_2016}, where they use the DEMNUni maps obtained via ray-tracing from $z_{\rm min}=0.02$ to $z_{\rm max}=20.88$. Here, we verify the presence of the same effects in the cross-spectra of DEMNUni maps obtained up to $z_{\rm max}=1.89$ and use these results to validate the analytical predictions computed using Equation~\eqref{ctp}.\\
The cross-correlations of the DEMNUni ISWRS and CMBL potential maps for the four neutrino masses are shown in the top panels of Figure~\ref{cross_TP}, compared with the theoretical nonlinear prediction computed for the \citetalias{Takahashi_2012} (top left panel of Figure~\ref{cross_TP}) and the \citetalias{Mead_2021} (top right panel of Figure~\ref{cross_TP}) models. For completeness, also the theoretical cross-correlation predictions for the massless neutrino case in the linear regime have been reported.

The agreement between the simulation data and both the theoretical predictions obtained with \texttt{CAMB} up to $\ell \sim 100$ is quite impressive.
Once exceeded $\ell \sim 100$, the anti-correlation between the RS effect and the CMBL potential takes over, resulting in the characteristic sign inversion that appears in both simulations (as already shown in~\cite{Carbone_2016}) and nonlinear predictions.
Moreover, both of them validate the expected result, because the more neutrinos are massive, the more they counteract nonlinear evolution (i.e. structure formation) and nonlinear effects appear at smaller cosmological scales. In other words, the more neutrinos are massive, the more the sign inversion shifts towards larger multipoles.  

The achieved results disprove the counter-intuitive effect reported in~\cite{Xu_2016}, where for larger $M_{\nu}$ values the sign inversion position moves towards smaller multipoles. The behaviour found in ~\cite{Xu_2016} is presumably due to the use of a very approximated implementation of the neutrino mass dependence in the Halofit models, which leads to different results from ours.

In the subpanel of Figure~\ref{cross_TP}, we focus on the sign inversion position as a function of  $M_{\nu}$, comparing the two \texttt{Halofit} models to DEMNUni. The $1\sigma$ error bars, shown for the DEMNUni measurements, have been estimated as follows: we generate 5000 Gaussian realisations of the DEMNUni maps and estimate the errors as the standard deviation. Further discussion will be provided in Section~\ref{sec:detection}.

\section{The ISWRS cross-correlation with galaxies}
\label{sec:XC-TG}
The cross-spectrum between the ISWRS signal and the galaxy distribution is the most used tool to investigate the ISWRS, because of its largest signal-to-noise ratio with respect to the spectra analysed in the previous Section. It is therefore the main focus of this work, which consists of verifying the presence of the sign inversion due to the anti-correlation between the RS effect and the galaxy distribution in the cross-correlation angular power spectra, showing that the position of this sign inversion depends on $M_{\nu}$, as in the case of the cross-correlation between the ISWRS signal and the CMBL potential~\cite{Carbone_2016}. 
The validation of these results via the DEMNUni maps represents an improvement with respect to previous works: on the one hand,~\cite{Lesgourgues_2008} computed this cross-correlation signal in massless and massive neutrino models, focusing only on the effects that neutrinos have in the linear regime; on the other hand, in the nonlinear regime,~\cite{Smith_2009} focused only on the massless neutrino case, finding in the cross-spectrum the sign inversion due to the anti-correlation between the RS effect and the galaxy distribution. 

\begin{figure*}[!ht]
\centering
\includegraphics[width=0.90\textwidth]{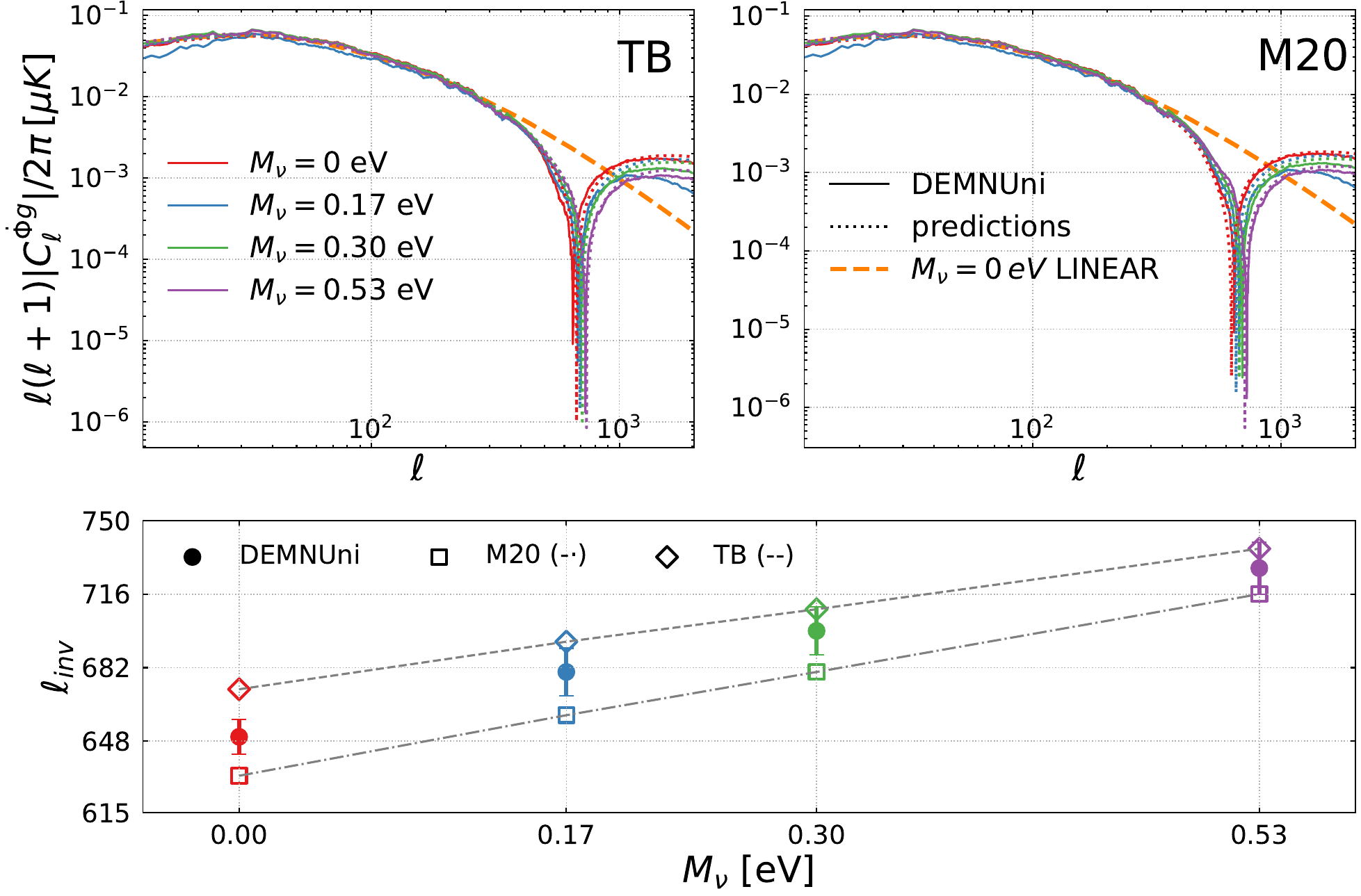}
\caption{ISWRS--galaxy cross-spectra for $M_{\nu} = 0,0.17,0.30,0.53\text{ eV}$ (solid red, blue, green and purple lines, respectively) compared with theoretical predictions in the linear (dashed orange line) and nonlinear (dotted lines) regimes, obtained using the \citetalias{Takahashi_2012} model (left) and the \citetalias{Mead_2021} model (right), respectively. The subpanel shows the behaviour of $\ell_{inv}$ as a function of $M_{\nu}$ for the two nonlinear models, compared with the measurements from the DEMNUni simulations.}
\label{cross_TG}
\end{figure*}

The cross-correlations of the DEMNUni ISWRS and the galaxy distribution maps for the four neutrino masses are shown in Figure~\ref{cross_TG}, compared with the theoretical nonlinear predictions, computed with Equation~\eqref{ctg}. For completeness, also the theoretical predictions for the massless neutrino case in the linear regime have been reported. Here, for a fair comparison against the DEMNUni data, the theoretical predictions have been computed using $n(z)$ in Equation~\eqref{eq:demnuni_n} and the $M_\nu$-dependent linear interpolation in $z$ of the measurements in Equation~\eqref{eq:bgg}.

Also in this case, because of the suppression massive neutrinos induce in the matter power spectrum, the larger $M_{\nu}$, the more nonlinearities appear on smaller scales and consequently the sign inversion position that characterises this cross-correlation is shifted towards larger multipoles.

Moreover, the comparison of the simulated spectra with the theoretical predictions for both linear and nonlinear regimes is again quite impressive at large cosmological scales, while 
for $\ell \gtrsim 100$ nonlinear spectra dramatically differ from the linear one, because of the presence of the sign inversion. This result validates again the analytical method against mocks and makes it an extremely powerful tool to predict cross-spectra of future surveys in both the linear and nonlinear regimes.

In the subpanel of Figure~\ref{cross_TG}, we focus on the sign inversion position, $\ell_{inv}$,  as a function of  $M_{\nu}$, comparing the two \texttt{Halofit} models to DEMNUni. The $1\sigma$ error bars, associated to the DEMNUni measurements, have been estimated as described in the previous Section. Further discussion will be provided in Section~\ref{sec:detection}.

\subsection{Nonlinear ISWRS--galaxy cross-spectra: effects of different bias modellings}
As already mentioned, differently from the galaxy selection function, the galaxy bias strongly depends on the neutrino mass. 
We perform several tests to evaluate how the use of different 
bias functions affects the comparison between predictions and the DEMNUni measurements. First, we produce theoretical predictions for the massive neutrino cases using both $b_{gg}(z, M_\nu>0$) and $b_{gg}(z, M_\nu=0$), and we verify that the more the neutrinos are massive, the more $b_{gg}(z, M_\nu=0$) fails in predicting the spectra (further details in Section~\ref{subsec:b1}). Second, we check, for the massless neutrino case alone, the impact of using a scale-dependent bias function for the theoretical predictions (further details in Section~\ref{subsec:b2}).
The results of these tests are shown in Figure~\ref{fig:table_bias} and  Figure~\ref{fig:bias_zcomparison_lcdm}, respectively. 

\subsubsection{\texorpdfstring\boldmath{}{}Massive neutrino impact on the scale-independent bias}
\label{subsec:b1}

\begin{figure*}[!ht]
\centering
 \includegraphics[width=0.9\textwidth]{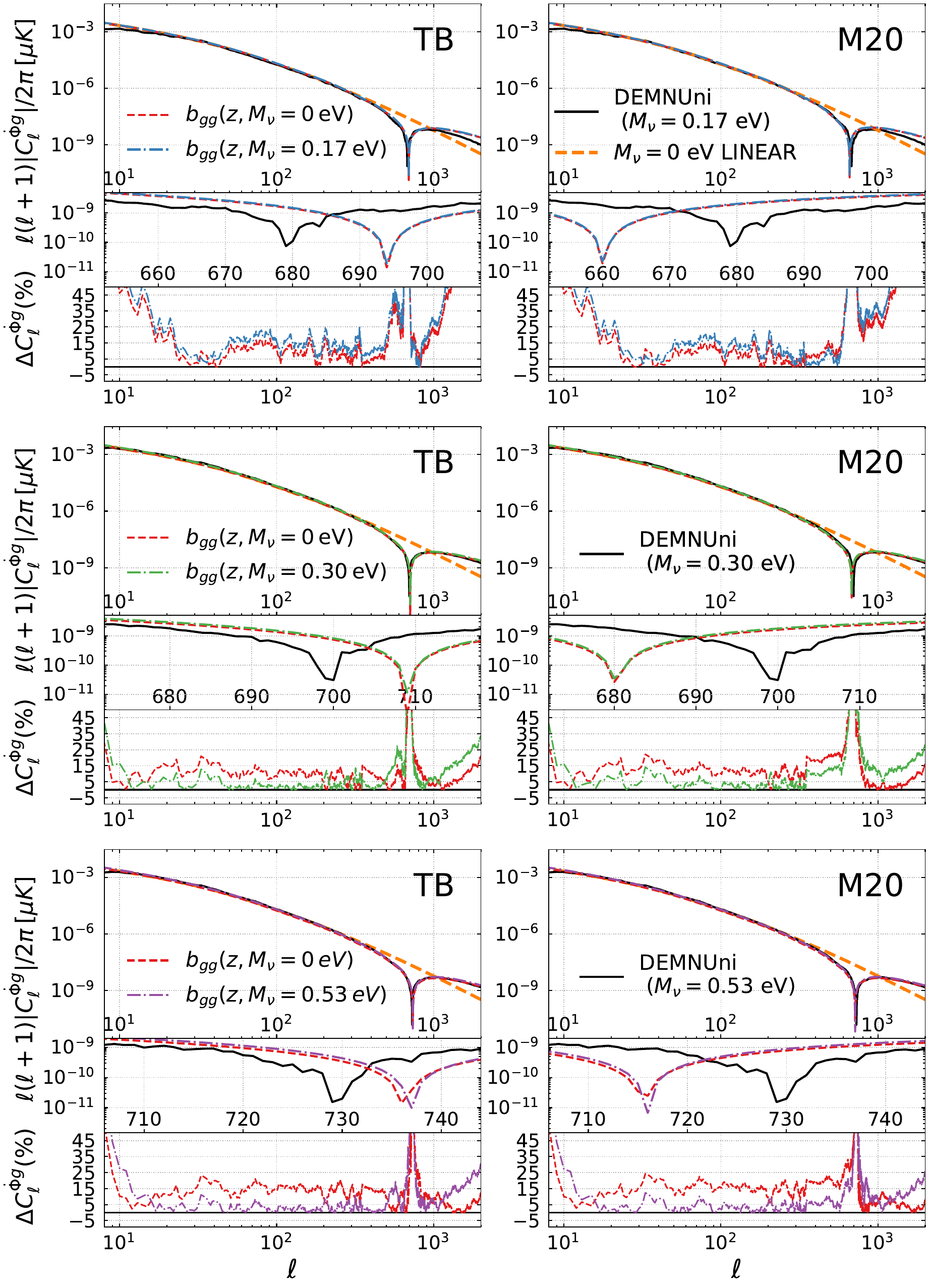}
\caption{Sensitivity to the $M_\nu$ dependence of the galaxy bias. \textit{Left}: Comparison between the simulated ISWRS--galaxy cross-correlations for $M_{\nu} = 0.17, 0.30, 0.53$ eV (solid black line) and theoretical predictions in the linear (dashed orange line) and nonlinear regimes, using the \citetalias{Takahashi_2012} model combined with DEMNUni measurements of $b_{gg}$ in the massless neutrino (dashed red) and  massive neutrino cases for $M_{\nu}=0.17\text{ eV}$ (\textit{top}, dot-dashed blue), $M_{\nu}=0.30\text{ eV}$ (\textit{middle}, dot-dashed green) and $M_{\nu}=0.53\text{ eV}$ (\textit{bottom}, dot-dashed purple), respectively. The middle subpanel of each plot shows a zoom-in of the sign inversion region, while the bottom one shows the absolute value of the percentage relative differences of the predictions with respect to the simulated measurements. \textit{Right}: Same as the left panel for the \citetalias{Mead_2021} model.}
 \label{fig:table_bias} 
\end{figure*}

First, we test how the analytical predictions for massive neutrino cosmologies compare with mock data when using the $b_{gg}(z,M_{\nu} > 0)$ bias function, or the $b_{gg}(z,M_{\nu}= 0)$.
Figure~\ref{fig:table_bias} shows such a comparison for the cross-correlation signals, for which the results for $\ell \le 10$ can be neglected because of finite volume effects in the mocks on those scales.

For the $M_{\nu} = 0.17$ eV case, differently from the other cosmologies, there is a small difference in the amplitude of the analytical cross-correlation power spectra, with respect to the DEMNUni ones, at very small ($\ell < 30$) and very large ($\ell > 1000$) multipoles. These differences were already visible in Figure~\ref{cross_TG}, and are evident in the top panels of Figure~\ref{fig:table_bias}. They induce a difference between the theoretical predictions and the DEMNUni cross-spectrum larger than $35\%$ for $\ell<30$ and $\ell>1000$. 

Before the sign inversion, for the \citetalias{Takahashi_2012} modelling the differences do not overcome $25\%$ for $b_{gg}(z , M_{\nu} = 0.17)$ and $15\%$ for $b_{gg}(z , M_{\nu} = 0)$. 
\clearpage
In the case of the \citetalias{Mead_2021} modelling, the difference between the predictions and measurements is slightly smaller ($< 20\%$) when using $b_{gg}(z, M_{\nu} = 0.17)$, 
and $\sim 15\%$ for $b_{gg}(z ,M_{\nu} = 0)$. 
The differences in the sign inversion positions are: $\Delta \ell_{\rm DEMNUni-(TB)} \sim - 18$ and $\Delta \ell_{\rm DEMNUni-M20} \sim + 18$. Concerning the signal amplitude, the predictions obtained with the two bias recipes show a difference at a level of $\sim 5\%$, while the sign inversions occur at the same multipole in the \citetalias{Mead_2021} model, and with a $\Delta \ell = 1$ for the \citetalias{Takahashi_2012} one. This result is somehow expected since the two scale-independent biases change only the overall amplitude of the signal but not its shape.
Instead, the fact that $b_{gg}(z, M_{\nu} = 0)$ in the theoretical predictions leads to smaller differences with respect to the DEMNUni spectra can be explained considering the inaccuracies of the \citetalias{Takahashi_2012} and \citetalias{Mead_2021} models in reproducing the $M_{\nu} = 0.17$ eV case, as already shown in Section~\ref{sec:theory}.

In the $M_{\nu}=0.30\text{ eV}$ case, shown in middle panels of  Figure~\ref{fig:table_bias}, both the nonlinear predictions, obtained with the \citetalias{Mead_2021} and \citetalias{Takahashi_2012} models, differ from the DEMNUni spectra by no more than $15\%$ when using $b_{gg}(z ,M_{\nu} = 0.30)$, and no more than $25\%$ when using $b_{gg}(z ,M_{\nu} = 0)$. The differences in the sign inversion positions are: $\Delta \ell_{\rm DEMNUni-(TB)} \sim -10$ and $\Delta \ell_{\rm DEMNUni-M20} \sim +20$. As expected, the difference between the two bias cases increases with the neutrino mass, specifically it is larger  by about $10\%$ than in the $M_\nu=0.17$ eV case. Concerning the sign inversions corresponding to the two biases, they occur at the same multipole in the \citetalias{Mead_2021} model, and with $\Delta \ell = 1$ for the \citetalias{Takahashi_2012} one.

Finally, as shown in bottom panels of Figure~\ref{fig:table_bias}, in agreement with the trend already noticed for $M_\nu=0.17, 0.30$ eV, for $M_{\nu}=0.53\text{ eV}$ the difference between the predictions using $b_{gg}(z ,M_{\nu} = 0.53)$ or $b_{gg}(z ,M_{\nu} = 0)$  increases up to $\sim 20\%$. The difference with respect to the DEMNUni spectra is $\sim 10\%$ 
for $b_{gg}(z ,M_{\nu} = 0.53)$, and $\sim 25\%$ for $b_{gg}(z ,M_{\nu} = 0$). Even in this case, the sign inversions occur at the same multipole when using \citetalias{Mead_2021}, and with $\Delta \ell = 2$  when using \citetalias{Takahashi_2012}. The differences in the sign inversion positions, with respect to the DEMNUni spectra, are: $\Delta \ell_{\rm DEMNUni-(TB)} \sim -10$ and $\Delta \ell_{\rm DEMNUni-M20} \sim +10$.

\subsubsection{\texorpdfstring\boldmath{}{}
\label{test_bias_mnu0}
Scale-independent and scale-dependent bias effects in the massless neutrino case}
\label{subsec:b2}
\begin{figure*}[!ht] 
\centering
 \includegraphics[width=0.90\textwidth]{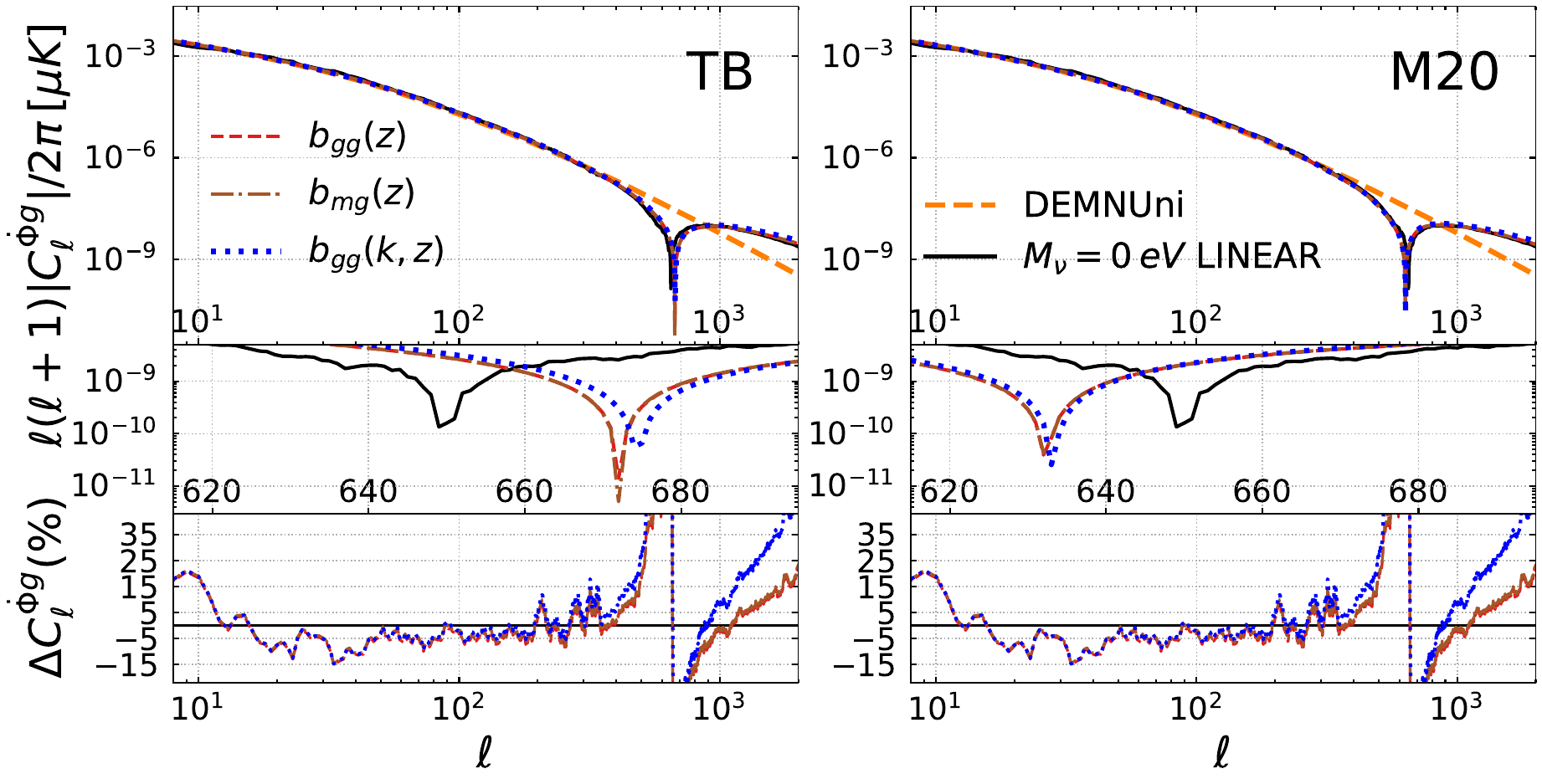} 
\caption{Sensitivity to the galaxy bias scale dependence. \textit{Left}: Comparison between the simulated ISWRS--galaxy cross-correlation for $M_{\nu}=0\text{ eV}$ (solid black line) with theoretical predictions in the linear regime (dashed orange line) and nonlinear regime
using the \citetalias{Takahashi_2012} model. Three different DEMNUni galaxy bias models have been tested: the scale-independent DEMNUni-$b_{gg}$ bias (dashed red), the scale-independent DEMNUni-$b_{mg}$ bias (dot-dashed brow) and the scale-dependent DEMNUni $b_{gg}(k,z)$  bias (dotted blue). The middle subpanel of the plot shows a zoom in of the sign inversion region, while the bottom one shows the absolute value of the percentage relative differences of the predictions with respect to the simulations. \textit{Right}: Same as left, but with the \citetalias{Mead_2021} model.}
\label{fig:bias_zcomparison_lcdm}
\end{figure*}

In Figure~\ref{fig:bias_zcomparison_lcdm} we compare the effects of $b_{mg}(z)$, $b_{gg}(k,z)$ and $b_{gg}(z)$ in the modelling of the theoretical nonlinear predictions.
Conversely to the work of  \cite{Smith_2009}, we find that the amplitude of the predictions obtained with the scale-dependent bias function does not show any particular difference with respect to the predictions computed with the scale-independent ones. The three reconstruction differ from DEMNUni spectra by no more than $\sim 10\%$.

With regard to the sign inversion position, we observe in Figure~\ref{fig:bias_zcomparison_lcdm} that,
when using the scale-independent prescriptions, independently of the \citetalias{Takahashi_2012} and \citetalias{Mead_2021} models, the sign inversions occur at mostly the same multipoles as for the scale-independent case for which $\Delta\ell_{\rm DEMNUni-TB} \sim -20$ and $\Delta \ell_{\rm DEMNUni-M20} \sim +15$, respectively.  In fact, the $\ell_{inv}$ obtained with $b_{gg}(k,z)$ differs by $\Delta_{\ell} = 2$  ($\Delta\ell_{\rm DEMNUni-TB} \sim -22$ ) with respect to the scale-independent prescriptions in the \citetalias{Takahashi_2012} case, and by $\Delta_{\ell} = 1$ ($\Delta\ell_{\rm DEMNUni-M20} \sim +14$) in the \citetalias{Mead_2021} one. We stress here that $M_\nu=0$ is the only case in which the \citetalias{Mead_2021} model works better than the \citetalias{Takahashi_2012} one.

From the results of this test, we deduce that the use of a scale-dependent or scale-independent bias is almost indifferent for our modelling purposes, and  consequently they are both valid prescriptions.

\subsection{Nonlinear ISWRS--galaxy cross-spectra:  the case of different galaxy surveys}
\label{sec:forecasts}
\begin{figure*}[!ht] 
\centering
\includegraphics[width=0.90\textwidth]{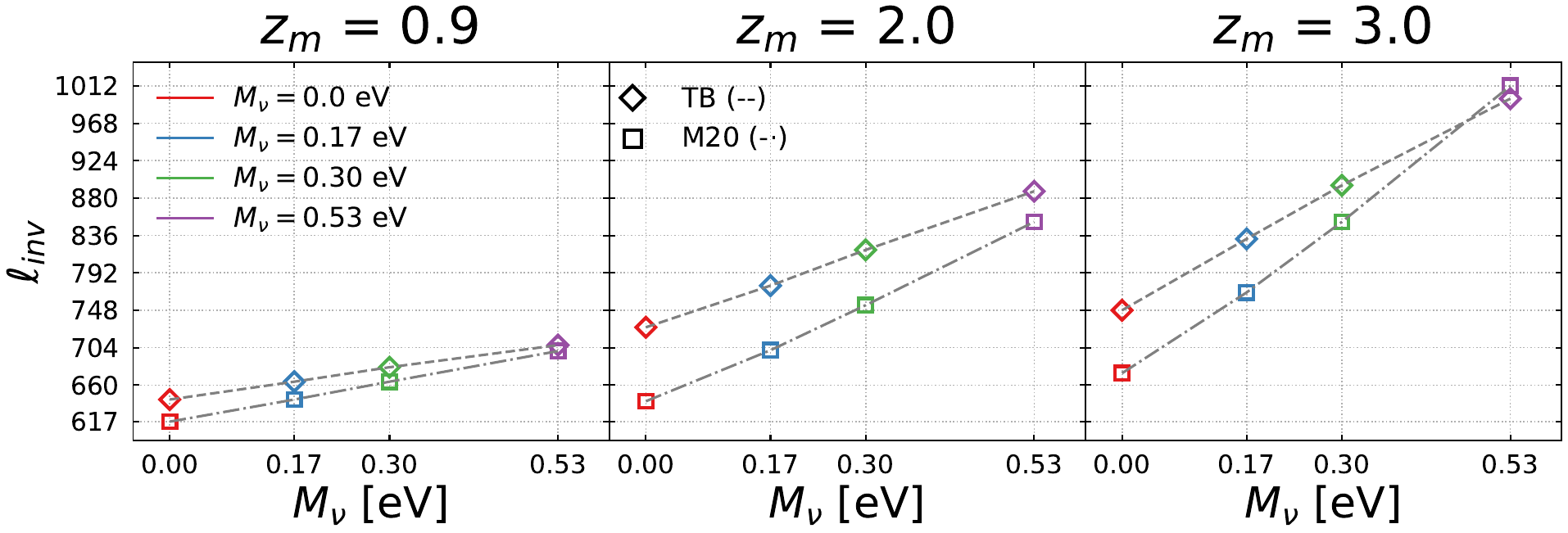} 
\caption{Forecast of the expected $\ell_{inv}$ as a function of $M_{\nu}$, computed using the three $n(z,z_{m})$ considered in this work, with $z_{m} = 0.9$ (\textit{left}),  $z_{m} = 2$ (\textit{center)} and  $z_{m} = 3$ (\textit{right}), testing both the \citetalias{Takahashi_2012} (diamonds)  and \citetalias{Mead_2021} (squares).}
\label{fig:table_forecast}
\end{figure*}

The last part of this work consists of computing predictions of the cross-spectra that can be observed on higher redshift ranges than the ones analysed so far, forecasting in particular their dependence on the parameter $z_{m}$, in Equation~\eqref{eq:Euclid_n}, that characterises a galaxy survey.
Here we compute the ISWRS--galaxy cross-spectra for the four $M_{\nu}$ values with three $n(z, z_{m})$ corresponding to $z_{m} = 0.9, 2, 3$.  To exploit the full redshift range where the $n(z, z_{m})$  are not zero, we compute our integrals up to $z_{max}=8$, conversely to the previous Sections where $z_{max}=1.89$ (i.e. the DEMNUni upper limit), and we use the $b^{Tut}_{gg}(z, M_{\nu})$ extended up to those redshifts via the fitting formula in Equation~\eqref{eq:euclid_bias}. To investigate the effect of the massive neutrinos presence, we show in Figure~\ref{fig:table_forecast} the behaviour of $\ell_{inv}$ as a function of $M_{\nu}$  for both the \texttt{Halofit} models.
It is evident at first glance that the higher the redshift, the larger is the difference between the four $M_{\nu}$ values, for both the \texttt{Halofit} models considered. This is because the late Universe (i.e. small $z$) is characterised by the presence of DE \cite{Giannantonio_2006}, that affects the appearance of nonlinearities and is expected to have similar effects as neutrinos on this cross-spectrum. Hence, the further we go back in time, the less DE effects will be present, and the shift we measure will be due only to the neutrino mass.

 \section{\texorpdfstring\boldmath{}{}Measuring the neutrino mass with nonlinear ISWRS-LSS cross-spectra?}
 \label{sec:detection}
The results shown in the previous Sections demonstrate the validity of the analytical method against the DEMNUni simulations and that both mocks and predictions confirm what we expect from theory: more massive neutrinos induce a larger suppression in the matter power spectrum and shift the appearance of nonlinearities on smaller cosmological scales.

The subpanels of Figures~\ref{cross_TP} and~\ref{cross_TG} show the behaviour of  $\ell_{inv}$ as a function of $M_{\nu}$ that we find in the two cross-correlation analyses. Both the nonlinear models have been tested and compared with the DEMNUni cross-spectra. 

First of all, we notice how in both cross-correlations and with both the \texttt{Halofit} models, the sign-inversion position, $\ell_{inv}$, varies almost linearly with $M_{\nu}$. Then we appreciate how the \citetalias{Takahashi_2012} model seems to work significantly better than the \citetalias{Mead_2021} one. In fact, when using \citetalias{Takahashi_2012}, for small values of the neutrino masses, we find an agreement with the DEMNUni simulations, at $\lsim 1\sigma$ level for the cross-correlation between the ISWRS and the CMBL, and at $\sim 1-1.5\sigma$ level for the cross-correlation with the galaxy distribution.  Conversely, the \citetalias{Mead_2021} model shows differences at $\gsim 2\sigma$ level for both cross-correlations.

\begin{figure*}[!ht]
\centering
\begin{tabular}{cc}
\includegraphics[width=0.48\textwidth]{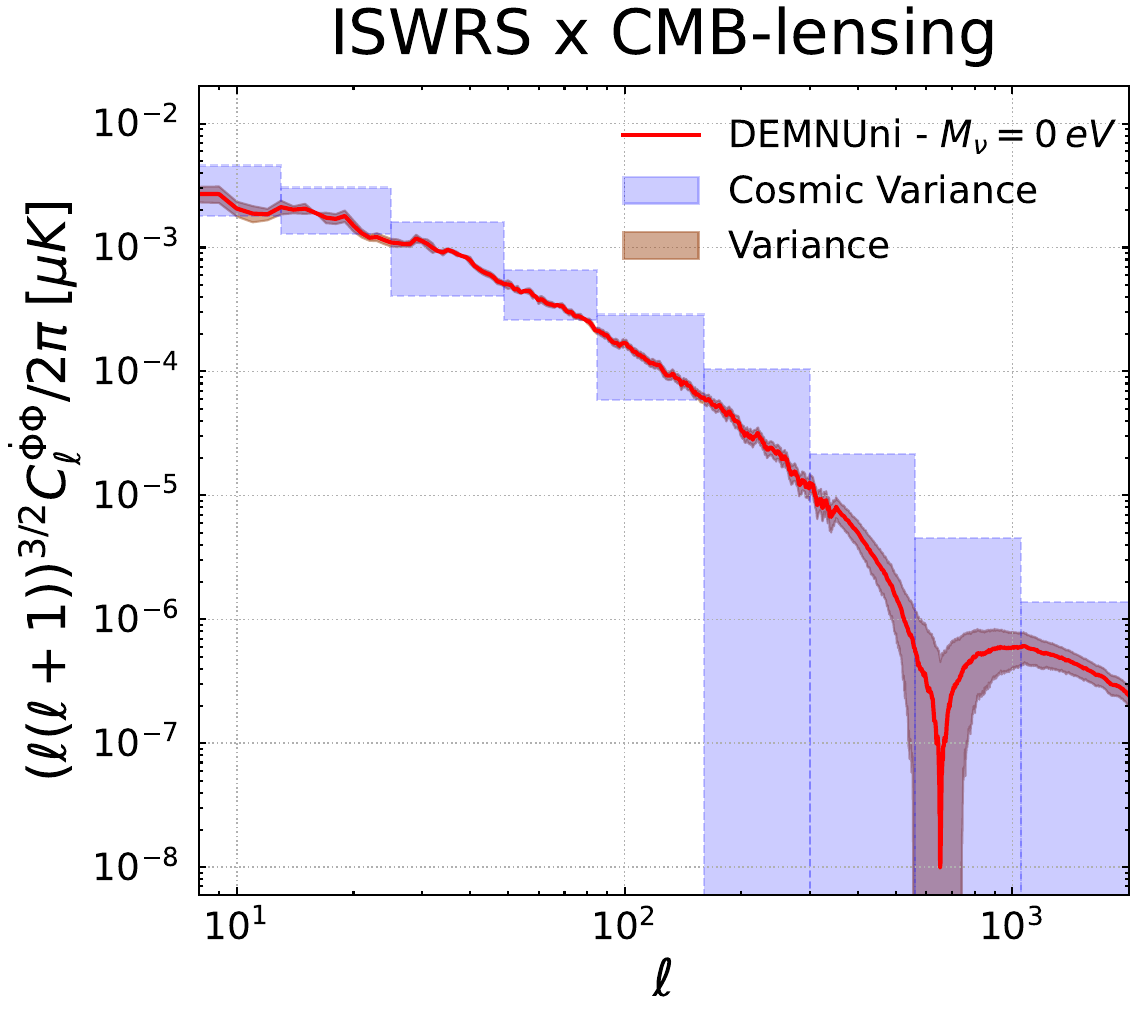}     &  \includegraphics[width = 0.48\textwidth]{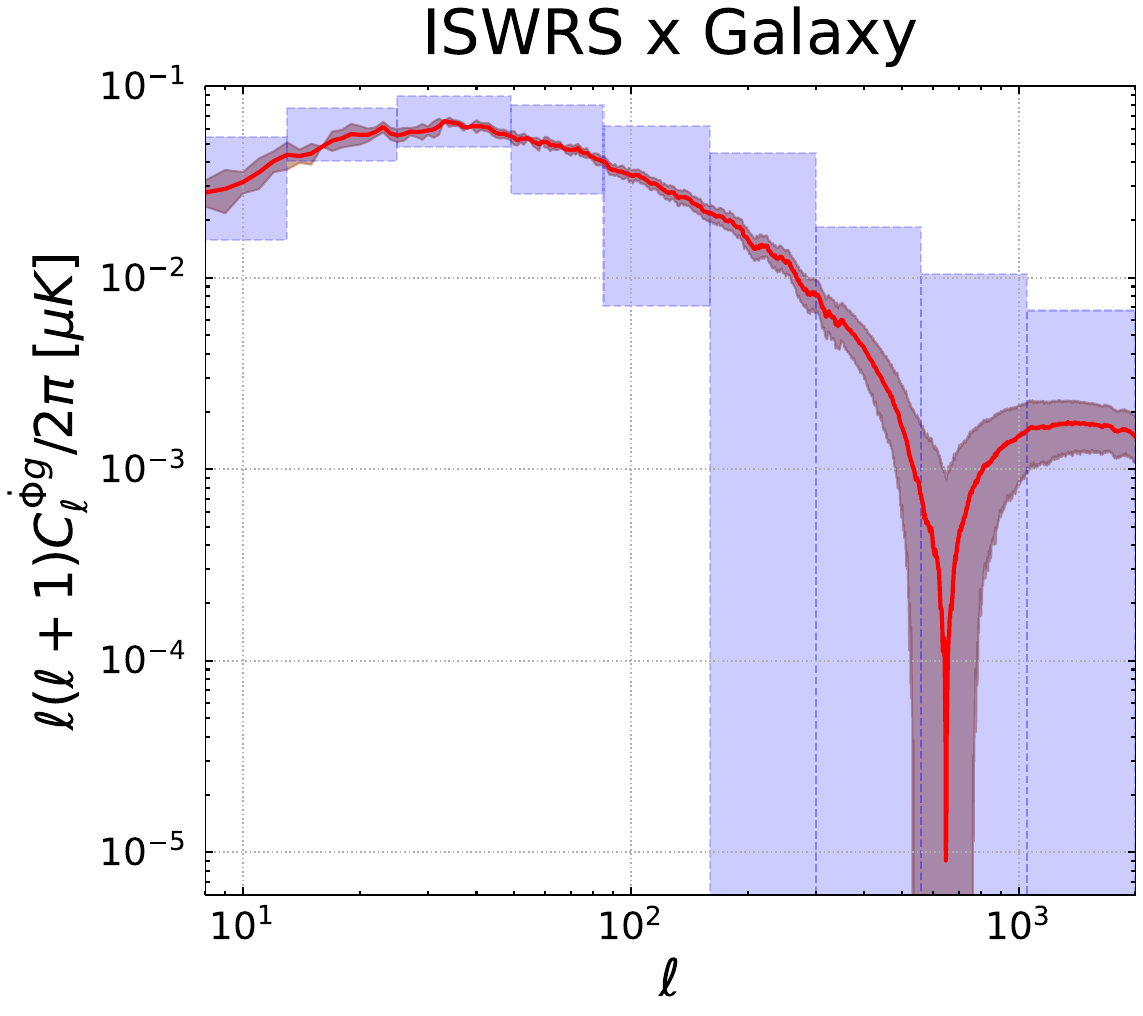}
\end{tabular}
\caption{\textit{Left}: ISWRS--CMBL potential cross-spectrum for $M_{\nu}= 0\text{ eV}$ (red line) with its cosmic variance binned in fourteen logarithmic equispaced intervals (shaded blue area) and the variance obtained as the standard deviation of the 5000 realisations computed (shaded brown area). \textit{Right:} Same as left, but for the ISWRS--galaxy cross-spectrum for $M_{\nu} = 0\text{ eV}$. In both panels, because of the choice to represent signals in range $\ell = [10,2000]$, only eleven bins are shown.}
\label{cosmic_variance}
\end{figure*}

Furthermore, for the massless neutrino case only, we compute the cosmic variance of both cross-spectra, as shown in Figure~\ref{cosmic_variance}. Already in an {\it idealised} cosmic variance limited case, it is evident that a detection of $\ell_{inv}$ is highly unlikely. Just for reference, in the same figure we also show the standard deviation of the $5000$ DEMNUni mocks. These represent the same sky realization, therefore they do not account for cosmic variance and their scatter is significantly smaller than the actual expected uncertainty on the cross-spectra. 

In principle, $\ell_{inv}$ could be sensitive even to the neutrino  hierarchy~\cite{Jimenez_2010,Lesgourgues_2012}.
However, current nonlinear modellings do not provide the required accuracy and precision. To our knowledge, nonlinear modellings of the matter power spectrum $P(k)$ for different neutrino hierarchies do not even exist yet. Furthermore, the difference in the matter power spectrum $P(k)$ between the two hierarchies is about $0.5 \%$ for a sum of neutrino masses of $0.1$ eV \cite{Wagner_2012}. Hence,  even in the case of future more accurate nonlinear prescriptions, measuring the shift of $\ell_{inv}$ due to neutrino hierarchies would be tremendously challenging.

Nevertheless, we might be able to disentangle different $M_{\nu}$ values by estimating the overall amplitude of the ISWRS signal on very small cosmological scales. This is because massive neutrinos suppress the $P(k)$ on small scales, resulting in an analogous suppression of the overall amplitude of the angular power spectra. Future CMB experiments, such as CMB-S4~\cite{CMB-S4}, will deliver high-resolution maps of wide portions of the sky, promising to allow for a detection of the ISWRS signal through the cross-correlation of CMB temperature maps with large-scale structure tracers. However, achieving a robust and statistically significant detection will require exquisite control of small scale extragalactic foregrounds in CMB maps, as emphasized in~\cite{Ferraro_2022}. Anyway, a comprehensive evaluation of the feasibility of such measurements and the corresponding constraints on $M_{\nu}$ demands a detailed analysis, currently part of an ongoing separate study.
 
\section{Conclusions}
\label{sec:conclusion}
The aim of this work is to develop and validate, against the DEMNUni cosmological simulations, an analytical modelling able to reproduce the angular cross-power spectra of the ISWRS signal with the CMBL potential and the galaxy distribution in the presence of massive neutrinos. 

We consider both the linear (ISW) and nonlinear (RS) effects, i.e. we account also for the nonlinear regime where a sign inversion appears in the cross-spectra due to the anti-correlation between the RS effect and the CMBL/galaxy distribution. We find that the presence of massive neutrinos affects the position of such a sign inversion, and specifically, the more massive neutrinos are, the more it shifts towards larger multipoles $\ell$. This is because hot free streaming neutrinos suppress the matter power spectrum, shifting the appearance of nonlinearities towards smaller cosmological scales.

There is already evidence of this effect at the level of the matter power spectra, $P_{\dot{\Phi}\Phi}$ and $P_{\dot{\Phi}\delta}$, associated to these cross-correlations, as we show in Figure~\ref{spectra_comp1_comp2}. We compute them assuming a Planck-like baseline cosmology, to which we add neutrinos with total mass $M_{\nu} = 0, 0.17, 0.30, 0.53 \text{ eV}$ and using two \texttt{Halofit} models implemented in \texttt{CAMB} (\citetalias{Takahashi_2012} and \citetalias{Mead_2021}) that allow us to investigate also high redshift ranges. We verify that a sign inversion appears in the transition from the linear to the nonlinear regime and that the presence of massive neutrinos affects its position. Moreover, as shown in Figure \ref{fig:table_forecast}, we find that the larger the redshift is, the smaller is the scale (the larger is the $\ell$)  where nonlinearities appear, because the RS effect dominates over the late ISW one, as we go back in time

Focussing on the nonlinear modelling, we find that \citetalias{Takahashi_2012} predicts a higher level of nonlinearities  with respect to \citetalias{Mead_2021} at small redshifts, but this trend is inverted at $z\ge  1.1$ (see Figure~\ref{spectra_comp1_comp2}).

The main result of this work is checking and modelling the presence of the same effects in the corresponding angular cross-power spectra, obtained from $P_{\dot{\Phi}\Phi}$ and $P_{\dot{\Phi}\delta}$, using Equations~\eqref{ctp}-\eqref{ctg}.

Concerning the cross-correlation between the ISWRS signal and the CMBL potential, we find that both the DEMNUni maps and the theoretical predictions, computed using the \citetalias{Takahashi_2012} and \citetalias{Mead_2021} models, show the sign inversion due to the anti-correlation between the RS and the CMBL potential, and the shift of this sign inversion moves towards larger multipoles as $M_{\nu}$ increases (see Figure~\ref{cross_TP}).
Furthermore, we observe that, while the \citetalias{Mead_2021} model tends to overestimate nonlinearities (and the sign inversions appear always on smaller multipoles with respect to the DEMNUni data), the \citetalias{Takahashi_2012} model gives different results depending on $M_{\nu}$, but the corresponding sign inversion locations are closer to the DEMNUni ones for all the $M_{\nu}$ values, and specifically overlap with the data for $M_{\nu}=0.17, 0.30$ eV. We conclude that in the case of the cross-correlation between the ISWRS effect and the CMBL potential, the \citetalias{Takahashi_2012} model works better than the \citetalias{Mead_2021} one in reproducing nonlinear effects and how they are affected by the presence of massive neutrinos.

We have performed a similar analysis for the cross-correlation between the ISWRS effect and the galaxy distribution  (see Figure~\ref{cross_TG}). This represents an improvement with respect to~\cite{Smith_2009}, where only the massless neutrino case was investigated, and extends the work of~\cite{Lesgourgues_2008} properly treating, in the nonlinear regime, the impact of massive neutrinos on this cross-correlation. In this case, we find that, as for the cross-correlation between the ISWRS effect and the CMBL potential, the \citetalias{Mead_2021} model again overestimates nonlinearities. Instead, for the \citetalias{Takahashi_2012} model, the trend here is to underestimate nonlinearities for all the $M_{\nu}$ values, but again the $\ell_{inv}$ are closer to DEMNUni ones with respect to the \citetalias{Mead_2021} case.

Considering the dependence on the galaxy bias of the cross-correlation between the ISWRS effect and the galaxy distribution, we have performed two further tests to quantify the effects of different bias modellings. First, we test how the analytical predictions fit the simulated signal when considering a bias (measured from the DEMNUni simulations) dependent or not on the neutrino mass. We observe that, for small values of $M_{\nu}$, the differences between the modelling with massless and massive neutrino bias are negligible (Figure~\ref{fig:table_bias}). Conversely, when considering large values of $M_{\nu}$, the difference cannot be neglected, and a bias model specific to that cosmology is required, as expected. Second, only in the massless neutrino case, we test the effect on the theoretical predictions of a scale-independent or scale-dependent bias models, both calibrated against measurements from the DEMNUni suite. We verify that, for both the \citetalias{Takahashi_2012} and the \citetalias{Mead_2021} modellings, choosing a scale-dependent bias $b_{gg}(k,z)$ produces analytical cross-spectra comparable to those obtained with a scale-independent bias $b_{gg}(z)$, as shown in Figure~\ref{fig:bias_zcomparison_lcdm}. Hence, we can conclude that both the bias models are valid for this kind of analysis.

Moreover, we have verified that the ISWRS-galaxy cross-spectra change with redshift because of the different role played by DE at different epochs. Therefore, a straightforward extension of this work would be to include DDE models in the analysis, and study the dependence of the ISWRS-LSS cross-spectra on the DE equation of state. This could help to distinguish between different DE models and provide insights on the properties and the evolution of DE.

Finally, our results highlight the strict connection between the sign inversion location, $\ell_{inv}$, and the value of neutrino masses. However, the cosmic variance limit on these measurements makes the detection of $\ell_{inv}$ highly unlikely. 
Nevertheless, the accuracy of future surveys might allow to distinguish different $M_{\nu}$ values via the estimation of the overall amplitude of the ISWRS signal and its cross-correlation with LSS on very small cosmological scales.

Overall in this work, we have developed an analytical modelling of the nonlinear ISWRS-LSS cross-correlations that exploits the Boltzmann solver code \texttt{CAMB} and has been validated against the DEMNUni simulations~\cite{Carbone_2016}. From the comparison of the two nonlinear \texttt{Halofit} models, we have observed that the \citetalias{Takahashi_2012} model works better for the reconstruction of the two cross-correlations analysed, with respect to the \citetalias{Mead_2021} one. Ours is a useful and computationally non-expensive tool to investigate these cross-correlations in different cosmological scenarios without the need to run large N-body simulations for each cosmology.
In a future work we will investigate the detectability of the neutrino mass via ISWRS-LSS cross-correlations. We expect future CMB experiments, such as The Simons Observatory~\cite{Lee_2019}, LiteBIRD\cite{LiteBird}, CMB-S4~\cite{Abazajian_2016}, and galaxy surveys such as \textit{Euclid}~\cite{EuclidI_2022}, the Roman Space telescope~\cite{Eifler_2021}, the Vera Rubin Observatory~\cite{Ivezic_2019}, will help us exploiting such nonlinear signals with the aim of measuring, in combination with more traditional probes, the neutrino mass and the DE EoS.

\acknowledgments 
VC and MM are partially supported by the INFN project ``InDark''. MM is also supported by the ASI/LiteBIRD grant n.2020-9-HH.0. MC is partially supported by the 2021/22 and 2022/23 ``Research and Education'' grant from Fondazione CRT. The OAVdA is managed by the Fondazione Cl\'ement Fillietroz-ONLUS, which is supported by the Regional Government of the Aosta Valley, the Town Municipality of Nus and the ``Unit\'e des Communes vald\^otaines Mont-\'Emilius''.
The DEMNUni simulations were carried out in the framework of ``The Dark Energy and Massive Neutrino Universe" project, using the Tier-0 IBM BG/Q Fermi machine and the Tier-0 Intel OmniPath Cluster Marconi-A1 of the Centro Interuniversitario del Nord-Est per il Calcolo Elettronico (CINECA). CC acknowledges a generous CPU and storage allocation by the Italian Super-Computing Resource Allocation (ISCRA) as well as from the coordination of the ``Accordo Quadro MoU per lo svolgimento di attività congiunta di ricerca Nuove frontiere in Astrofisica: HPC e Data Exploration di nuova generazione'', together with storage from INFN-CNAF and INAF-IA2.

\bibliographystyle{JHEP}
\bibliography{bibliography.bib}

\appendix

\end{document}